\documentclass[preprint]{aastex}
\usepackage{natbib,epsfig,wrapfig,ulem}

\def\h2o{H$_2$O}
\def\ch4{CH$_4$}
\def\arcs{\ifmmode {''}\else $''$\fi}

\shorttitle{Characterizing Very Low Mass Companions}
\shortauthors{Rice et al.}

\begin{document}

\title{A New Method for Characterizing Very-Low-Mass Companions with Low Resolution Near-Infrared Spectroscopy}

\author{Emily L.\@~Rice\altaffilmark{1,2}, Rebecca~Oppenheimer\altaffilmark{2}, Neil Zimmerman\altaffilmark{3}, Lewis C.\@~Roberts, Jr.\altaffilmark{4}, Sasha Hinkley\altaffilmark{5,6}}

\altaffiltext{1}{Department of Engineering Science and Physics, College of Staten Island, City University of New York, Staten Island, NY 10314, USA, emily.rice@csi.cuny.edu}
\altaffiltext{2}{Department of Astrophysics, American Museum of Natural History, New York, NY 10024, USA}
\altaffiltext{3}{Department of Mechanical and Aerospace Engineering, Princeton University, Princeton, NJ 08544, USA}
\altaffiltext{4}{Jet Propulsion Laboratory, California Institute of Technology, Pasadena CA 91109, USA}
\altaffiltext{5}{Astronomy Department, California Institute of Technology, Pasadena, CA 91125, USA}
\altaffiltext{6}{Department of Physics \& Astronomy, University of Exeter, Devon EX4~4QL, UK}

\begin{abstract}
We present a new and computationally efficient method for characterizing very low mass companions using low resolution ($R\sim$30) near-infrared ($YJH$) spectra from high contrast imaging campaigns with integral field spectrograph (IFS) units. We conduct a detailed quantitative comparison of the efficacy of this method through tests on simulated data comparable in spectral coverage and resolution to the currently operating direct imaging systems around the world. In particular, we simulate Project 1640 data as an example of the use, accuracy, and precision of this technique. We present results from comparing simulated spectra of  M, L, and T dwarfs with a large and finely-sampled grid of synthetic spectra using Markov Chain Monte Carlo techniques. We determine the precision and accuracy of effective temperature and surface gravity inferred from fits to {\tt PHOENIX} {\it dusty} and {\it cond}, which we find reproduce the low-resolution spectra of all objects within the adopted flux uncertainties. Uncertainties in effective temperature decrease from $\pm$100--500~K for M dwarfs to as small as $\pm$30~K for some L and T spectral types. Surface gravity is constrained to within 0.2--0.4~dex for mid-L through T dwarfs, but uncertainties are as large as 1.0~dex or more for M dwarfs. Results for effective temperature from low-resolution $YJH$ spectra generally match predictions from published spectral type-temperature relationships except for L-T transition objects and young objects. Single-band spectra (i.e., narrower wavelength coverage) result in larger uncertainties and often discrepant results, suggesting that high contrast IFS observing campaigns can compensate for low spectral resolution by expanding the wavelength coverage for reliable characterization of detected companions. We find that SNR$\sim$10 is sufficient to characterize temperature and gravity as well as possible given the model grid. Most relevant for direct-imaging campaigns targeting young primary stars is our finding that low-resolution near-infrared spectra of known young objects, compared to field objects of the same spectral type, result in similar best fit surface gravities but lower effective temperatures, highlighting the need for better observational and theoretical understanding of the entangled effects of temperature, gravity, and dust on near-infrared spectra in cool low gravity atmospheres.
\end{abstract}


\section{\sc Introduction}
Extra-solar planets have been indirectly detected for over 20 years, and there are more nearly 2,000 known systems, with over twice as many {\it Kepler} candidates awaiting confirmation.\footnote{http://exoplanetarchive.ipac.caltech.edu/} Indirect detection techniques can provide mass, radius, and bulk composition measurements for extra-solar planets, but they are limited in their utility to derive atmospheric properties, which are essential to testing formation scenarios \citep[e.g.,][]{Fortney08,Spiegel12}. Recent direct imaging of extra-solar planets \citep[e.g.,][]{Kalas08,Marois08,Marois10,Lagrange10,Currie14} and studies of transiting planets in secondary eclipse \citep[e.g.,][]{Deming07,Knutson08,Sing09,Fortney13} have expanded the field to include pseudo-spectral studies via photometric observations made at multiple wavelengths, including narrow band filters \citep{Janson10}. 

Integral field spectroscopy will rapidly expand the study of substellar companion and exoplanetary atmospheres. In the last few years, direct detection and spectroscopy has revealed a diversity of atmospheric properties, even within the same planetary system \citep{Patience10,Bowler10,Barman11a,Konopacky13,Oppenheimer13,Chilcote15}. Several ground-based high-contrast imaging instruments have been specifically designed to directly image extra-solar planets, including: the Near-Infrared Coronagraphic Imager (NICI) on Gemini-South \citep{Chun08}, the High Contrast Instrument for the Subaru Next Generation Adaptive Optics (HiCIAO) and later the Coronagraphic High Angular Resolution Imaging Spectrograph on (CHARIS) on Subaru \citep{Suzuki10,McElwain12}, Project 1640 (P1640) on the Palomar Observatory Hale 200-inch telescope \citep{Hinkley11,Oppenheimer12}, the Gemini Planet Imager (GPI) on Gemini-South \citep{Macintosh14}, and the Spectro-Polarimetric High-contrast Exoplanet Research (SPHERE) on the VLT \citep{Beuzit08}. 

P1640, GPI, and SPHERE include integral field spectrograph (IFS) units that produce low resolution spectra (spectral sampling $R$=[$\lambda$/$\Delta\lambda$]$\sim$30--60) of directly imaged companions in the near-infrared ($\sim$1--2~$\mu$m). The high contrast ratios achieved by these instruments greatly expands the parameter space of the companions for which spectroscopy can be obtained, including M, L, \& T dwarf and planetary companions around BAFG stars \citep{Zimmerman10,Hinkley10,Hinkley11b,Roberts12,Hinkley13,Chilcote15,Crepp15}, and simultaneous observations of multiple-planet systems \citep{Oppenheimer13}, all at Solar System-like separations (4--40~AU). The low resolution of spectra obtained by these instruments are crucial for confirming the nature of the companions and for constraining their atmospheric properties and formation scenarios.

The primary purpose of IFS units on high contrast instruments is to distinguish faint companions from residual light from the primary star that has been diffracted through the instrument in a complex pattern \citep{Oppenheimer09}. The IFS exploits the wavelength dependence of the diffraction in order to separate speckles from faint objects, which will remain at the same position on the image plane while the speckles change position with wavelength \citep{Sparks02}. The spectral sampling and coverage of IFS units is motivated by speckle suppression rather than spectral characterization of detected companions and limited by the engineering complexities of the lenslet arrays and the pixel and chip size of infrared detectors. 

Very low-mass stars, brown dwarfs, and planetary-mass companions have potential to be efficiently characterized by comparing observed low resolution spectra to synthetic spectra from model atmospheres, as long as the selected wavelength regime contains significantly broad spectral features that are sensitive to atmospheric properties like effective temperature, surface gravity, and metallicity. Spectral fitting thus avoids assumptions about age, radius, and mass, as well as the evolutionary models that relate these quantities, e.g., \citet{Burrows97,Chabrier00,Baraffe02,Fortney07}. As higher-contrast and lower-mass objects are imaged and observed spectroscopically, the ability to conduct such comparisons efficiently and reliably is needed. 

We have developed a method to derive atmospheric parameters and uncertainties by fully mapping a broad temperature-gravity parameter space using thousands of synthetic spectra and linearly interpolating between calculated models. We compare simulated P1640 spectra (described in Section~\ref{obssim}) of field M, L, and T dwarfs and young M and L spectral type objects to synthetic spectra calculated with the {\tt PHOENIX} model atmosphere code (Section~\ref{phx}) using a Markov Chain Monte Carlo (MCMC)-based fitting procedure described in Section~\ref{fits}. Section~\ref{results} presents the resulting probability distribution functions and best fit model spectra, describes the constraints on parameters of effective temperature and surface gravity, compares results obtained using different subsets of the simulated spectra, and discusses the dependence of best fit parameters and uncertainties on the noise level in the simulated spectra. The conclusions of this work are presented in Section~\ref{conclu}.

\section{\sc Spectral Templates and Simulated Spectra}
\label{obssim}
In order to test our fitting procedure over a range of effective temperatures (hence companion masses), we constructed low resolution ($\Delta\lambda$$\sim$300~\AA) spectra from moderate-resolution, high signal-to-noise ratio (SNR) observed spectra of field M, L, and T dwarfs and young M and L dwarfs. The original observational spectra are henceforth referred to as ``template" spectra and the constructed lower-resolution spectra as ``simulated"  spectra. The creation of simulated low-resolution near-infrared spectra from the template spectra is described below.

\subsection{Spectral Templates}
The sample of objects used as spectral templates is summarized in Table~\ref{obs}. Their spectral types range from M1 to T4.5, spanning the range of effective temperatures expected for low mass companions that can be detected by the current generation of high contrast integral field spectrographs \citep{Beichman10}. All template spectra were obtained with the SpeX spectrometer on the NASA Infrared Telescope Facility (IRTF, \citealt{Rayner03}) in cross-dispersed mode, covering  0.81--2.4~$\mu$m with a resolving power ($R$=$\lambda$/$\Delta\lambda$) of at least 2000, with the exception of the spectrum of PSO J318.5338$-$22.8603, which was obtained on the GNIRS spectrograph \citep{Elias06} on the Gemini-North 8.1-m Telescope in cross-dispersed mode, covering 0.95--2.5 $\mu$m with $R\sim$1700.

High-contrast direct-imaging surveys will primarily target young stars for which low-mass companions are likely to be still contracting, thus they have lower surface gravity than field-age objects of the same mass. Template objects include both field (ages $>$1~Gyr) M, L, and T dwarfs and young ($\sim$10--100~Myr) M and L dwarfs. Spectra of field objects are from the IRTF Spectral Library \citep{Cushing05,Rayner09}. The young objects are an M8.5$\gamma$ \citep{RFC10}, an L0$\gamma$ \citep{Kirkpatrick06}, an L5$\gamma$ (\citealt{Faherty13}, see \citealt{Cruz09} for complete description of the $\gamma$ gravity suffix), and an L7 {\sc vl-g} \citep{Allers13,Liu13}. These young objects are high probability members of $\beta$~Pictoris, Tucana-Horologium, AB~Doradus, and $\beta$~Pictoris, respectively (\citealt{RFC10,Gagne14,Faherty13,Liu13}) and are among the closest photometric and spectroscopic analogs of directly-imaged gas giant planets for which detailed analysis is possible (see, e.g., \citealt{Faherty13,Allers13}).

\subsection{Simulated High Contrast IFS Spectra}
\label{P1640sim}
We simulate low-resolution near-infrared spectra of very-low-mass stellar and substellar companions that can be obtained by high contrast IFS units in order to test the accuracy and precision of physical parameters inferred from model fits to these spectra. We report results for spectra representative of the resolution and wavelength range for Project~1640, which has a single observing mode covering the $YJH$ band with a sampling of $\Delta\lambda$$\sim$300~\AA. 

Project~1640 has operated since 2008 on the Hale telescope in two distinct phases, Phase~1 and Phase~2, separated by a substantial instrument upgrade (new detector, filters, and increased sensitivity, \citealt{Oppenheimer12}). Phase~1 (July 2008 to June 2012) produced $JH$ (1.1--1.8~$\mu$m) spectra at 300~\AA~pixel$^{-1}$ resolution with one instrument setting \citep[see][]{Hinkley08,Hinkley11}. Phase~2 upgrades (begun June 2012) extended the wavelength coverage to $YJH$ (0.9--1.8~$\mu$m) and slightly increased the spectral sampling to 286~\AA~pixel$^{-1}$ \citep{Oppenheimer12,Oppenheimer13}. Simulated P1640 spectra were created from SpeX/IRTF template spectra for both phases, but the results of this paper focus on fits to Phase~2 spectra.

\begin{figure}
  \includegraphics[height=.75\textheight,angle=90,origin=c]{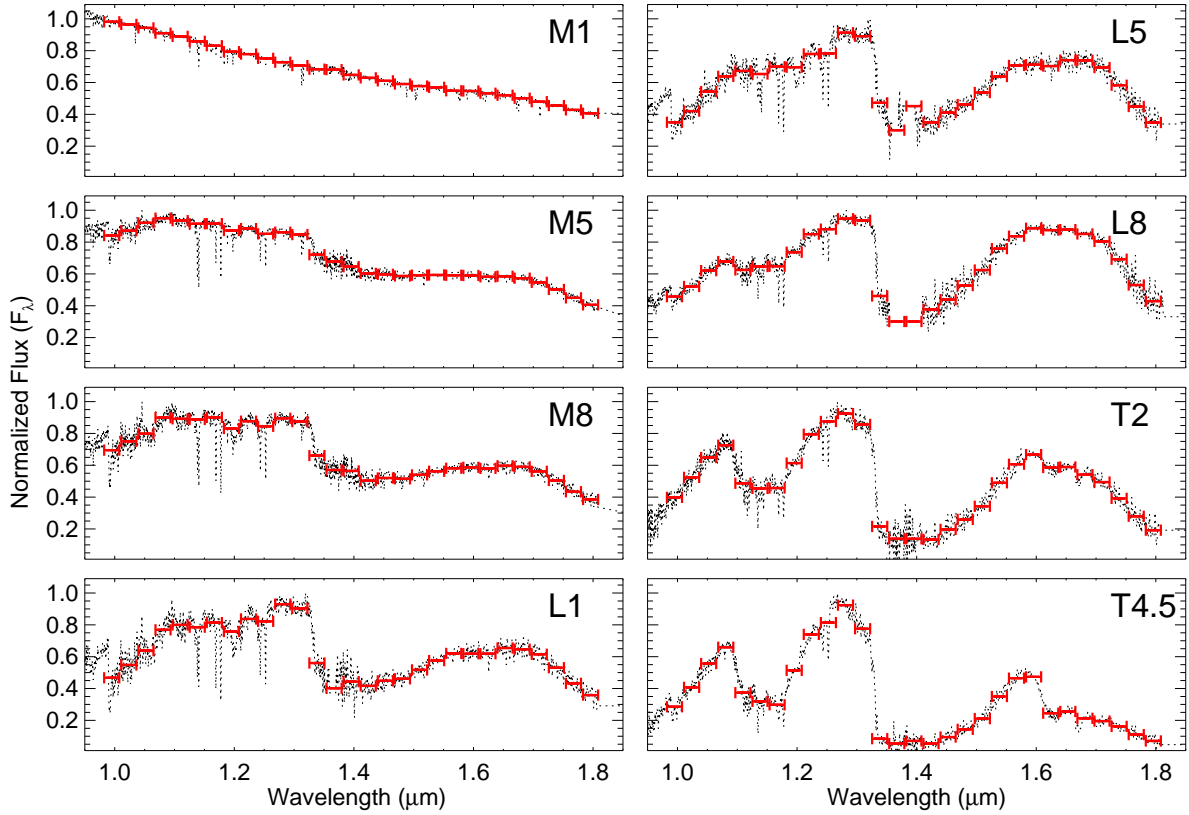}
  \caption{\label{IRTF_P1640}Observed (dashed lines) and simulated (red bars) near-infrared spectra of field M, L, and T dwarfs from IRTF/SpeX library spectra \citep{Cushing05,Rayner09}. The binned spectra simulate the wavelength coverage and spectral sampling of P1640 $YJH$ spectra (Phase~2). The calculation of the simulated spectra is described in Section~\ref{P1640sim}. The spectral template sources are listed in Table~\ref{obs}.
}
\end{figure}

\begin{figure}
  \includegraphics[height=.75\textheight,angle=90,origin=c]{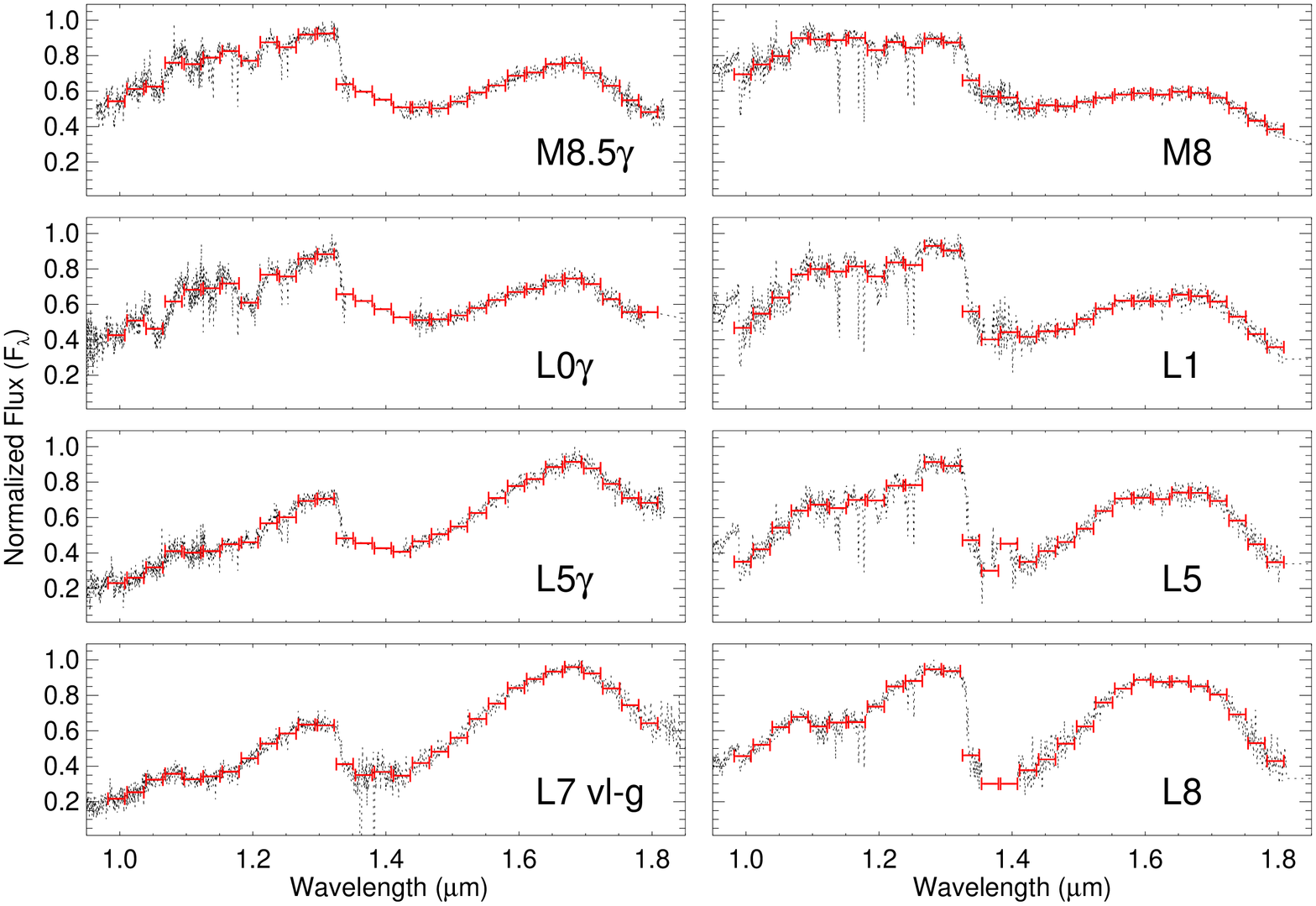}
  \caption{\label{IRTF_P1640_young}Observed (dashed lines) and simulated (red bars) near-infrared spectra of young M and L spectral type objects (left; \citealt{RFC10,Kirkpatrick06,Faherty13,Allers13,Liu13}) and their closest field-age counterparts in our sample (right). The binned spectra simulate the wavelength coverage and spectral sampling of P1640 $YJH$ spectra. The calculation of the simulated spectra is described in Section~\ref{P1640sim}. The spectral template sources are listed in Table~\ref{obs}.
}
\end{figure}

The first step in the creation of simulated P1640 spectra is to define the wavelength channels based on the data reduction and spectral extraction procedures developed by the instrument team \citep{Zimmerman11,Oppenheimer13}. At each wavelength channel, the flux density ($F_{\lambda}$) value of the simulated spectrum was set to the median of the flux density from the template spectrum within the wavelength range of the channel. The quoted resolution of the binned spectra represents the spectral sampling rather than the full-width at half-maximum of the instrumental profile. The simulated spectra were then normalized to the maximum value in the P1640 wavelength range. The template IRTF spectra and simulated P1640 spectra are plotted in Figures~\ref{IRTF_P1640} and \ref{IRTF_P1640_young}. 

The simulated spectra are significantly higher SNR than the typical spectra of very-low-mass companions obtained with ground-based high contrast integral field spectrographs \citep[e.g.,][]{Oppenheimer13,Chilcote15}. Therefore, for the primary analysis described in Section~\ref{fits}, the uncertainty on the simulated spectra was set to SNR=10 at the peak flux value and assigned a constant noise at each flux point, resulting in lower SNR for lower flux values. The actual uncertainties on these flux points (i.e., errors on the template spectra flux added in quadrature within each wavelength channel) the are much smaller; therefore, we test the dependence of best fit parameters and uncertainties on SNR via several ``resampling" tests described in Section~\ref{snr}.

The simulated spectra do not explicitly include any instrumental systematics such as speckles or cross-talk \citep[see][]{Oppenheimer09}. The effects of residual uncorrected wavefront aberrations, or speckles, coinciding with a detected companion in the extracted image cube is prohibitively time-consuming to predict and incorporate into the our analysis, although they are minimized with current data reduction and spectral extraction procedures \citep{Pueyo12,Oppenheimer13,Fergus14}. Another potentially significant instrumental effect is cross-talk, or light from different wavelengths leaking in to neighboring spectra on the detector. For Phase~2 data, cross-talk has been reduced to a maximum of 1.5\% and average of 0.4\% across all wavelengths with better focusing of the instrument, an improved blocking filter, and optimized extraction methods \citep{Oppenheimer13}.

\section{\sc Model Atmospheres and Synthetic Spectra}
\label{phx}
Model atmospheres and synthetic spectra were calculated using the {\tt PHOENIX} atmosphere code \citep{Hauschildt97,Hauschildt99}. For an effective temperature range 1400--4500~K, appropriate for M and L dwarfs, the {\it dusty} version of the {\tt PHOENIX} code was used, as presented in \citet{Rice10}, \citet{Schlieder12}, and \citet{Roberts12}. The {\it cond} version of the {\tt PHOENIX} code was used for an effective temperature range of 950--2000~K, appropriate for mid to late L dwarfs and early to mid T dwarfs as presented in \citet{Rice10aas} and similar to \citet{delBurgo09}. 

The {\it dusty} and {\it cond} versions of the {\tt PHOENIX}  models represent limiting cases of dust treatments. In the  {\it dusty} version, dust is created in thermo-chemical equilibrium, assumed to be made of spherical single-species grains with a power-law size distribution, and remains static in the atmosphere as a source of opacity. In the  {\it cond} version, dust is created identically but the opacity is not considered in calculation of the emergent spectrum, thus mimicking perfectly efficient removal of the dust to layers below the photosphere. These limiting cases have been shown to work well for objects hotter than mid-M type \citep[e.g.,][]{Rice10} and cooler than the L-T transition \citep[e.g.,][]{delBurgo09,Rice10aas}. While more sophisticated dust treatments have been implemented in other atmosphere models, we use {\tt PHOENIX} models because we have access to a complete and finely sampled model grid over the required range of effective temperature and surface gravity.

To test the veracity of our {\tt PHOENIX} {\it dusty} and {\it cond} synthetic spectra compared to low-resolution near-infrared spectra of M, L, and T dwarfs, we added a nuisance parameter to the fitting procedure described in Section~\ref{fits} below. This parameter is added in quadrature to the actual uncertainty on the binned flux points (i.e., not the adopted uncertainty for SNR$\le$10) and allowed to vary. The T$_{eff}$ and log(g) results from these fits marginalized over the tolerance parameter were within the 1-$\sigma$ uncertainties of the reported results for all spectra and the nuisance parameter was always less than 0.10, with typical peaks around 5\%. This indicates that the systematic differences between the models and the simulated spectra are $\sim$5\% and as little as 1-2\% for the field M dwarfs.

For the model grids, atmospheric temperature-pressure structures were calculated in intervals of 50~K and with surface gravities ranging from log(g)=3.0--6.0 [cgs] in 0.1~dex intervals, resulting in 1952 {\it dusty} models and 682 {\it cond} models. We include surface gravities up to log(g)=6.0 (some unphysical) in our model grid in order to test the inherent sensitivity of the spectra to surface gravity in the absence of independent age constraints and to avoid a possibly artificial upper limit at log(g)=5.5. This also enables more reliable uncertainty estimates at high surface gravities because the probability distributions are not restricted to the physically allowed values. All atmosphere structures were calculated at solar metallicity. 

\begin{figure}
  \includegraphics[height=.75\textheight,angle=90,origin=c]{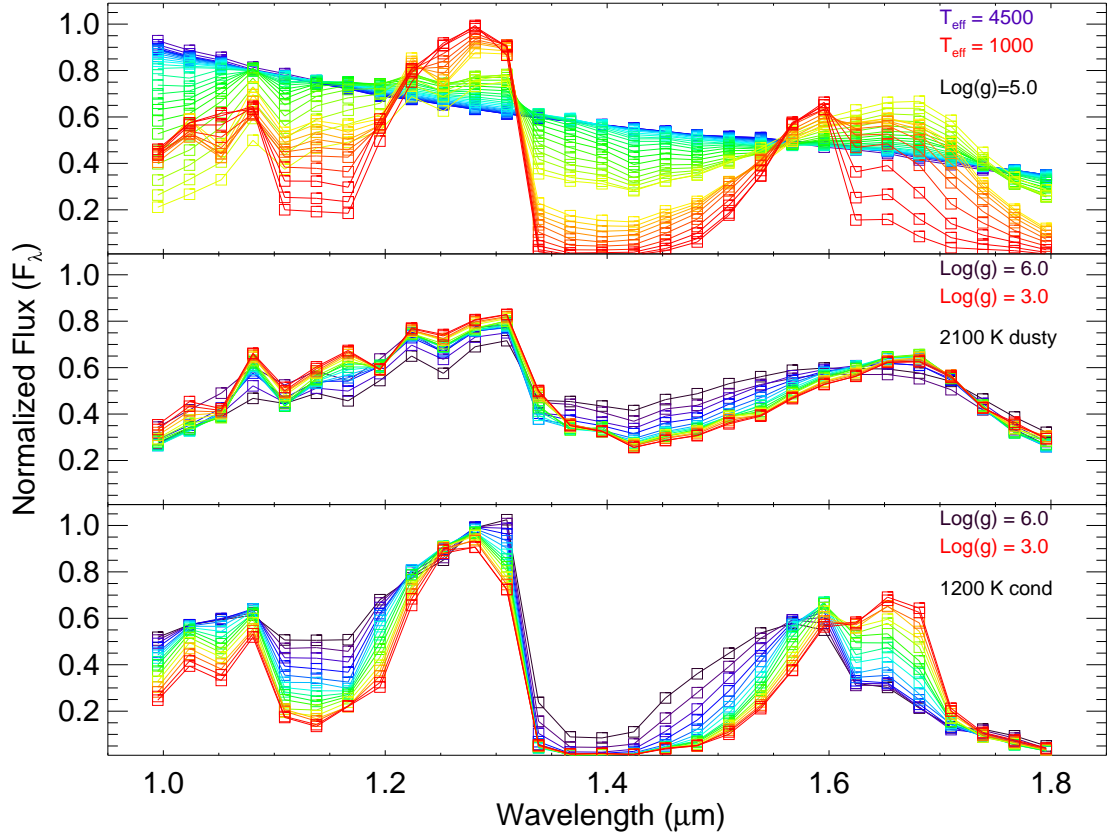}
  \caption{\label{lowresmodels}Synthetic spectra calculated with {\tt PHOENIX} model atmosphere code trimmed and binned to match P1640 spectra as described in Section~\ref{phx}. The top panel shows spectra for the complete range of model temperatures in 100~K increments at a fixed surface gravity log(g)=5.0 [cgs] expected for field objects.  The {\it dusty} models are shown for a fixed $T_{eff}\ge$2000~K, and {\it cond} models are shown for $T_{eff}\le$1900~K, hence the gap between spectra apparent at $\sim$1.4~$\mu$m. The middle panel shows spectra for the range of surface gravities at 2100~K (approximately late-M/early-L spectral types) from {\it dusty} models, and the bottom panel shows the same at a fixed $T_{eff}$=1200~K (approximately early-mid T dwarfs) from {\it cond} models.
}
\end{figure}

The calculation of a converged atmospheric structure with the {\tt PHOENIX} code also produces a synthetic spectrum at 4~\AA~sampling from 10~\AA~to 5~$\mu$m; therefore, the models oversample the simulated P1640 spectra by a factor of $\sim$70. Synthetic spectra matching the wavelength range and spectral sampling of P1640 are created following the same method used for simulated spectra, described in Section~\ref{P1640sim} above. Binned synthetic spectra for the range of effective temperatures and surface gravities used in the fitting procedure are shown in Figure~\ref{lowresmodels}. Within the model fitting procedure (described in Section~\ref{fits}) the calculated synthetic spectra are first linearly interpolated in flux to produce spectra with the desired temperature and gravity values, then binned to match the wavelength coverage and spectral sampling of the simulated P1640 spectra as described in Section~\ref{P1640sim}.

\section{\sc Spectral Fitting Method} 
\label{fits}
We use a two-step spectral fitting procedure to derive best fit physical parameters and uncertainties. The fitting procedure uses a grid of model spectra that are $\sim$10 times higher spectral sampling than the comparison (i.e., observed or simulated) spectrum and calculated at fine enough sampling in parameter space to allow reliable linear interpolation in flux (as described in \citealt{Rice10} and below). The linear interpolation is done on the fly by the fitting code, effectively mimicking a continuous grid of model spectra in the specified parameter space. In the current analysis we consider model parameters of effective temperature and surface gravity. Future work will include super-solar abundances expected for extra-solar planets created via core accretion \citep[e.g.,][]{Fortney08,Oberg11}. The fitting procedure is derived from the methods of \citet{Rice10} and updated in \citet{Roberts12}, \citet{Hinkley13}, and \citet{Crepp15}.

\begin{figure}
  \includegraphics[height=0.9\textheight,angle=90,origin=c]{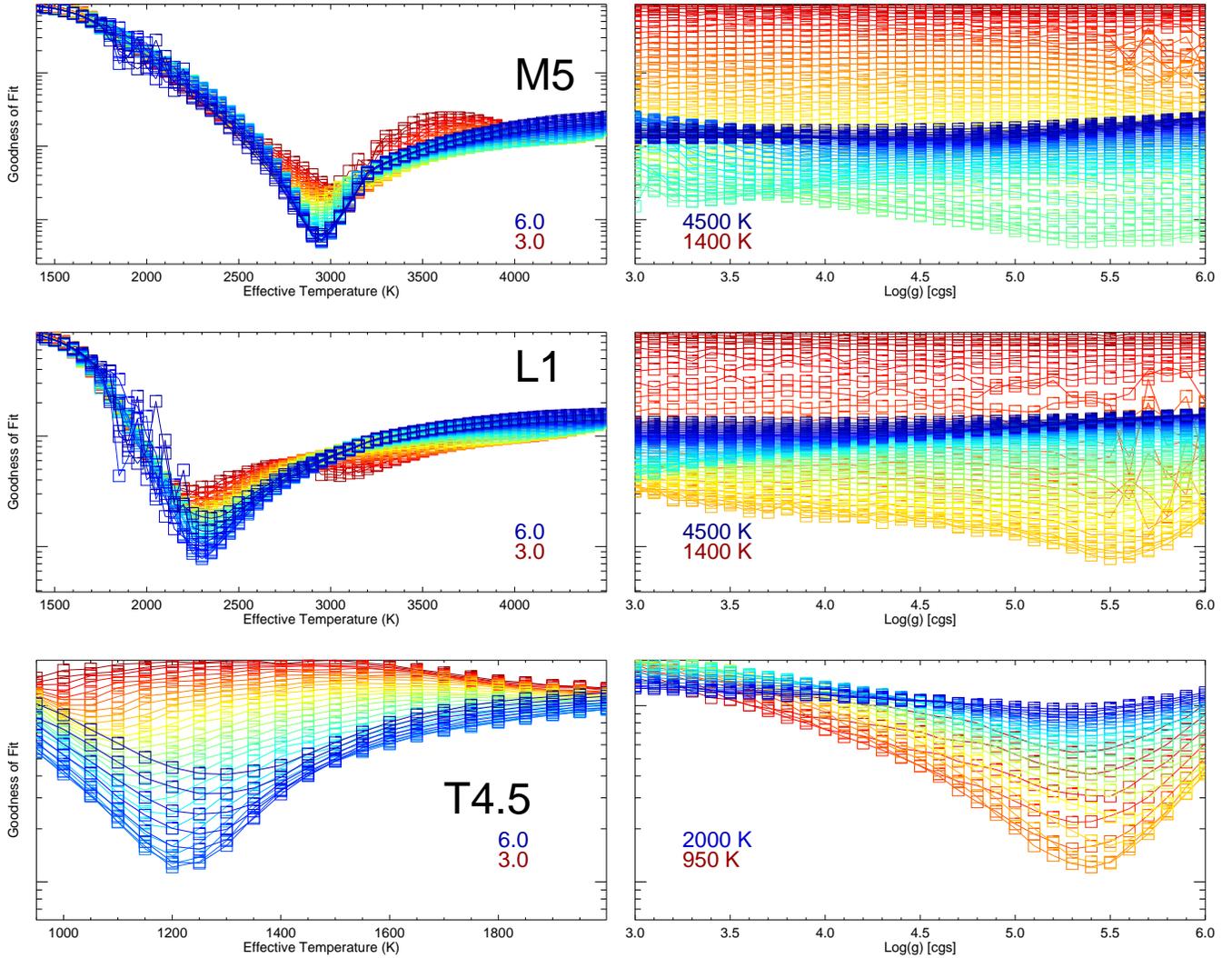}
  \caption{\label{goodnessoffit}Goodness of fit values for model spectra as a function of effective temperature (left) and surface gravity (right) for the simulated P1640 $YJH$ spectra from the M5 (top), L1 (middle), and T4.5 (bottom) spectral templates. This represents the first step in the fitting process and provides an overview of the complete parameter space. In these examples the lowest goodness of fit values are for model parameters 2950~K and logg=5.3 [cgs] for the M5 template, 2300~K and logg=5.5 [cgs] for the L1 and 1200~K and logg=5.3 [cgs] for the T4.5. These values are used as starting points for the MCMC procedure.The overview of parameter space provided by this step in the fitting procedure demonstrates that the low resolution near-infrared spectra of very low mass objects are sensitive to temperature (as expected) and are increasingly sensitive to surface gravity for later spectral types.
}
\end{figure}

The first step of the procedure uses the calculated model grids described in Section~\ref{phx}. For each synthetic spectrum we calculate a goodness-of-fit value based on the weighted sum of squared errors (see \citealt{Cushing08}). This calculation provides a coarse overview of the entire parameter range. Example results of this step are presented in Figure~\ref{goodnessoffit} for the $YJH$ simulated spectra of the field M5, L1, and T4.5 objects. 

The second step of the procedure uses a Metropolis-Hastings-based MCMC method to derive the posterior probability distribution over the model parameters. We assume an uninformative prior, i.e., one that is uniform over the range of parameters in the model grid. The uniform prior ignores prior knowledge about the objects that would make certain parameter regions higher or lower probability. We use it in order to present the most broadly applicable results and uncertainties without incorporating potentially inaccurate prior assumptions. The chain is initialized using the parameters of the model with the lowest (best) goodness of fit values from the previous step. Subsequent models are generated on the fly via linear interpolation of flux points from the pre-calculated grid in order to analyze an effectively continuous grid of models. We tested the veracity of linear interpolation in temperature by comparing spectra created via linear interpolation between two models 100~K apart to the calculated model with the same temperature. The mean difference for each flux point was 0.7\% for the {\it dusty} models and 1.4\% for the {\it cond} models, each with $\sigma$$\sim$2\% so we expect the systematic differences from linear interpolation to be at most 1--2\%. 

The uncertainties on the simulated or observed spectra are used to calculate the $\chi^2$ for the samples drawn from the posterior distribution. The ``jump size'' (width of the normal distribution from which the new model parameters are randomly selected) was adjusted for each object to produce an acceptance rate (number of accepted jumps divided by the total number of links) of $\sim$0.3--0.4. The jump sizes varied from 12.5 to 1450~K in effective temperature and 0.025 to 2.9~dex in surface gravity, with larger jump sizes for earlier spectral types. If a jump took the parameter outside the calculated model grid, then the interpolation step sets the parameter to just inside the grid.

The MCMC chain length for each spectrum is 10$^6$ links, where each link is a sample from the posterior distribution, whether or not the new sample is accepted into the chain. Typically in MCMC analysis the first $\sim$10\% of the links are considered ``burn-in" and not included in the final analysis; however, initializing the chain with the parameters with the lowest global goodness of fit values effectively removes the need for a burn in period. We confirm this by comparing the mean and standard deviations of the first $10^5$ steps in the chain with the 9$\times$10$^5$ post-burn links, which were nearly identical. Therefore the complete 10$^6$-link chains are used in analysis presented in Section~\ref{results}.

For preliminary tests of the fitting procedure, spectra were fit with the both {\it dusty} and the {\it cond} models with the appropriate temperature ranges for each dust treatment (described in Section~\ref{phx}). This, as expected, resulted in poor fits to the M~dwarfs using the {\it cond} models and to the T~dwarfs using the {\it dusty} models. For the L~dwarfs, the {\it dusty} model fits resulted in qualitatively better fits to the data; therefore, only the {\it dusty} models were used for the complete analysis of the L~dwarfs. Preliminary tests were also run with fewer links (10$^4$-10$^5$) with similar best fit results, indicating that the 10$^6$-link chains are converged.

P1640 spectra include the \h2o absorption band at $\sim$1.4~$\mu$m that is present in objects $\sim$M4 and later as well as in the Earth's atmosphere. Because the signal is low and the correction for telluric absorption is often poor, this region is typically excluded from analysis for observed spectra \citep[e.g.,][]{Roberts12,Oppenheimer13}. Therefore, we also fit ``trimmed" simulated P1640 spectra by excluding four flux points in the $\sim$1.4~$\mu$m \h2o absorption band. GPI and SPHERE cover the $YJHK$ spectral range with several individual filters (e.g., $Y$, $J$, $H$, $K1$, and $K2$ for GPI, \citealt{Macintosh08}). Therefore we fit the simulated P1640 spectra as separate $YJ$-band and $H$-band spectra. Results of these fits are discussed in Section~\ref{trim} below. Future papers will present this analysis for the exact wavelength coverage and spectral sampling of GPI and SPHERE.

\section{\sc Results}
\label{results}
\subsection{Probability Distributions}
\label{pdfs}
\begin{figure}
  \includegraphics[height=.75\textheight,angle=0]{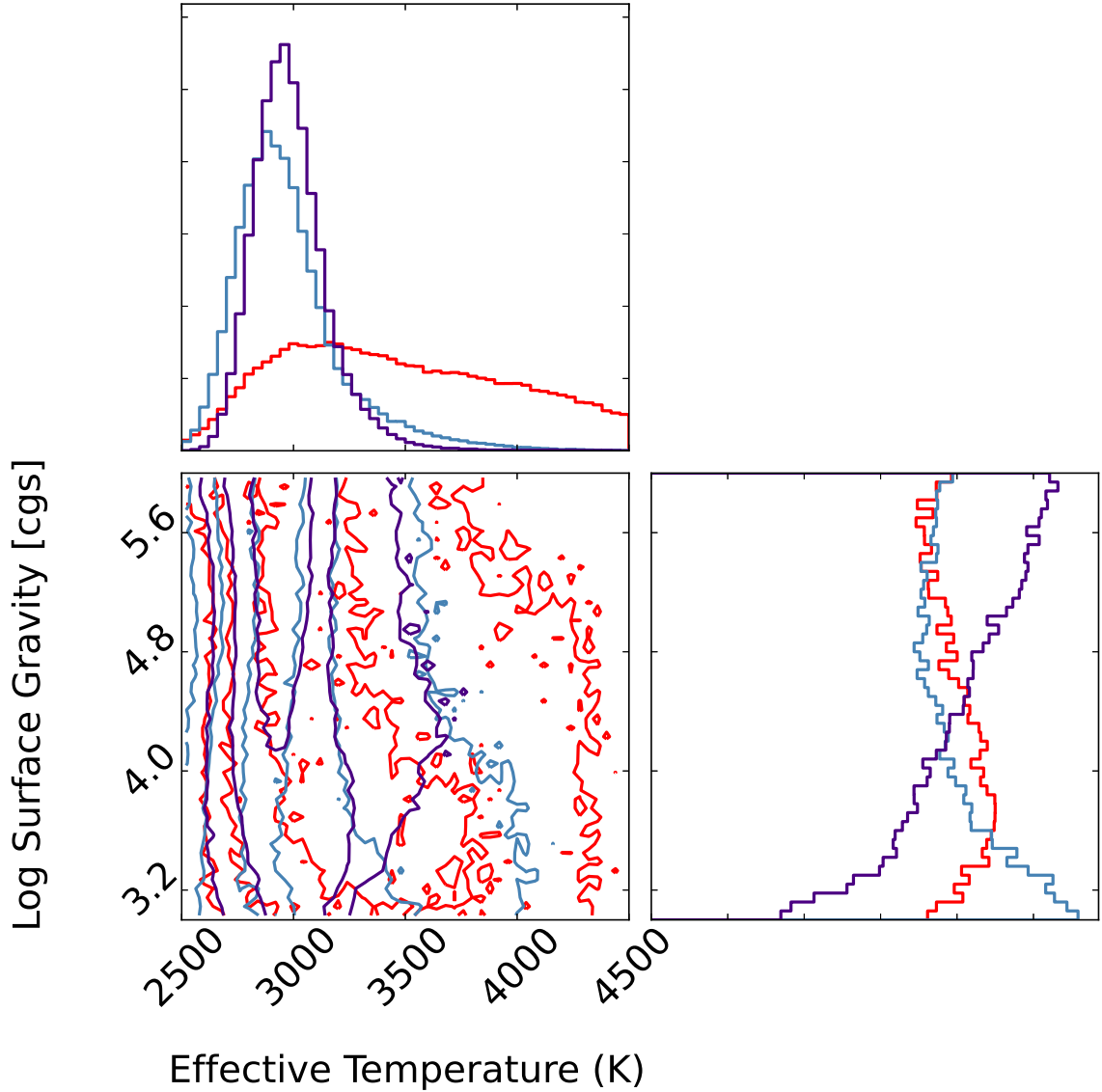}
  \caption{\label{mcmc_contours_M5}MCMC results for the field M5 spectral template object for: complete $YJH$ spectrum (purple), $YJ$ spectrum (blue) and $H$ spectrum (red). The lower-left panel shows 1-, 2-, and 3-$\sigma$ contours of the posterior distribution functions for both parameters, while the top and right panels shows the histograms for temperature and surface gravity, respectively, marginalized over the other parameter. The distributions for the $YJH$ and $YJ$ spectra are much narrower in temperature than in surface gravity. The $YJH$ and $YJ$-band spectra produce similar temperature results, but the $H$-band distribution has an only slightly sloping tail toward higher temperatures. 
}
\end{figure}

\begin{figure}
  \includegraphics[height=.75\textheight,angle=0]{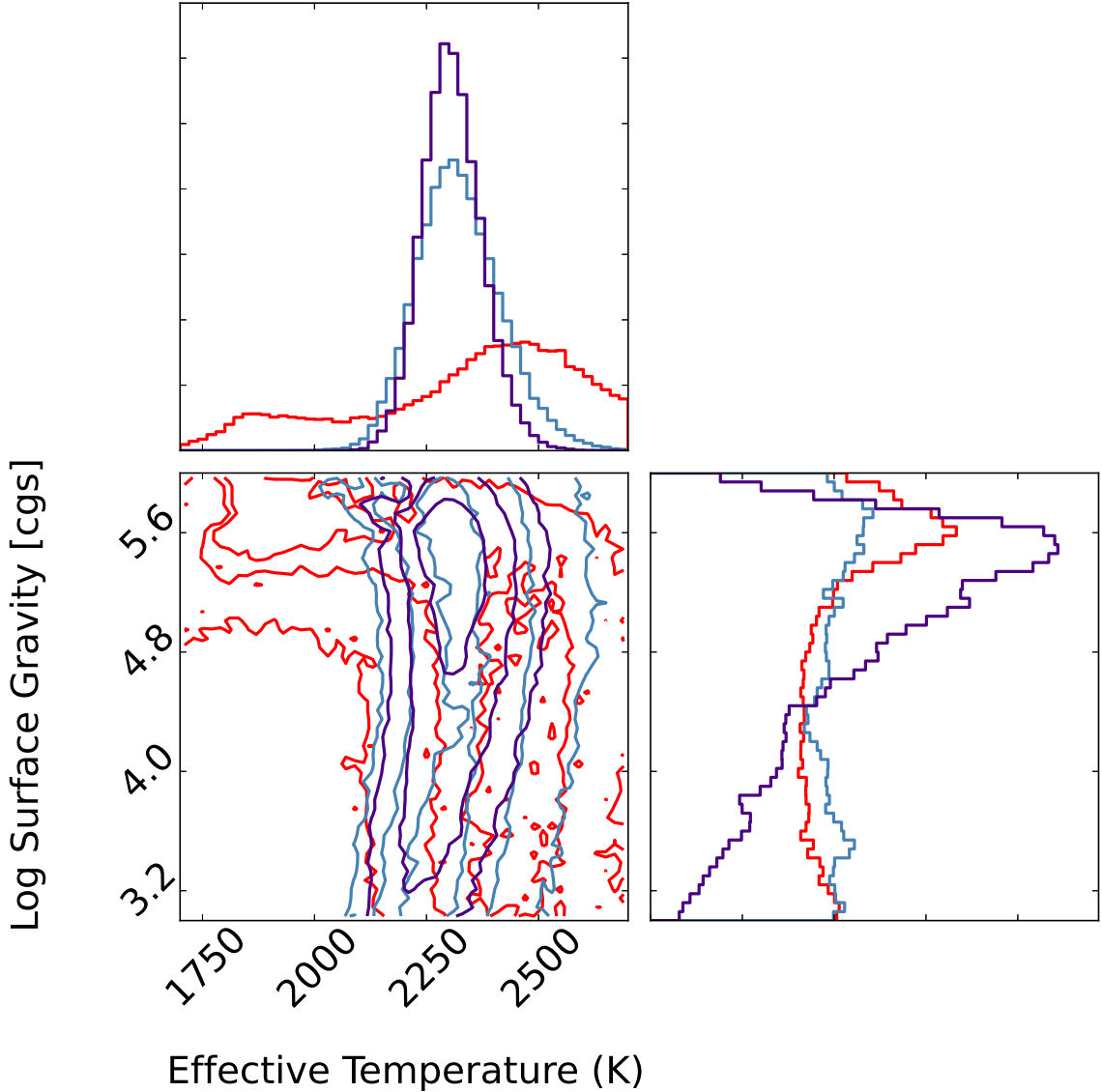}
  \caption{\label{mcmc_contours_L1}Same as Figure~\ref{mcmc_contours_M5} for the field L1 spectral template object. The results for the complete $YJH$ spectrum are better constrained in temperature than in surface gravity. Temperature is symmetric, while gravity has a longer tail to lower values. $YJ$-band spectra result in a more precise temperature estimate, while $H$-band and the complete $YJH$ spectrum provide a more reliable gravity estimate as indicated by the sharp decrease in probability at gravities higher than $\sim$5.6, which are unphysically high but allowed in the model grid, and the less pronounced decrease in probability for log(g)$<$5.0.
}
\end{figure}

\begin{figure}
  \includegraphics[height=.75\textheight,angle=0]{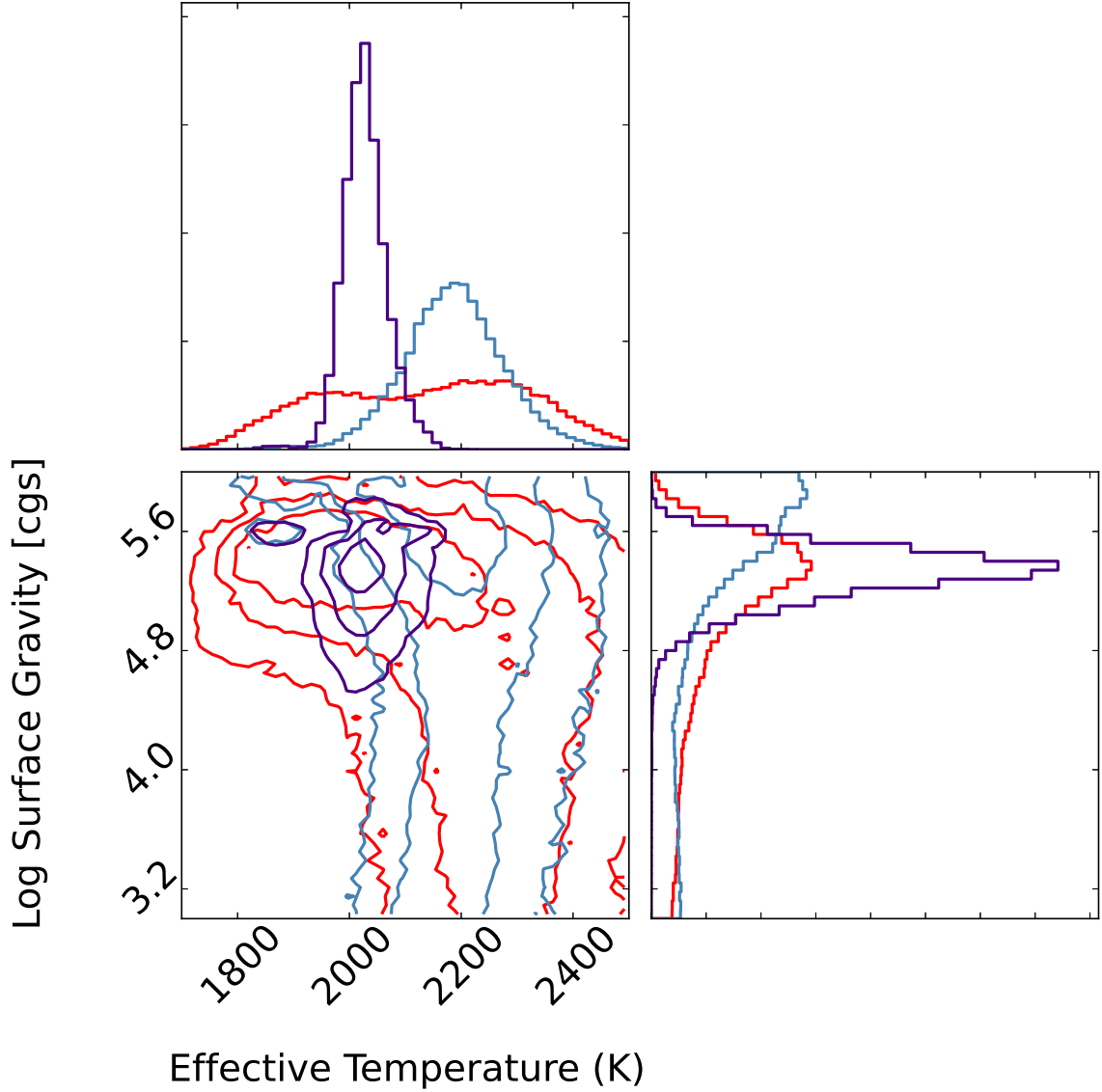}
  \caption{\label{mcmc_contours_L8}Same as Figure~\ref{mcmc_contours_M5} for the field L8 spectral template object. The distribution for the $YJH$ spectrum is narrow and symmetric in both temperature and gravity. The $YJ$distribution has a broader, but still single-peaked, distribution in temperature but increases toward unphysical high values in surface temperature. The $H$-band distribution is broad in temperature, covering the entire range of both the $YJH$ and $YJ$ distributions in temperature. In gravity the $H$-band distribution is consistent with the $YJH$ distribution with a long, flat tail toward lower values.
}
\end{figure}

\begin{figure}
  \includegraphics[height=.75\textheight,angle=0]{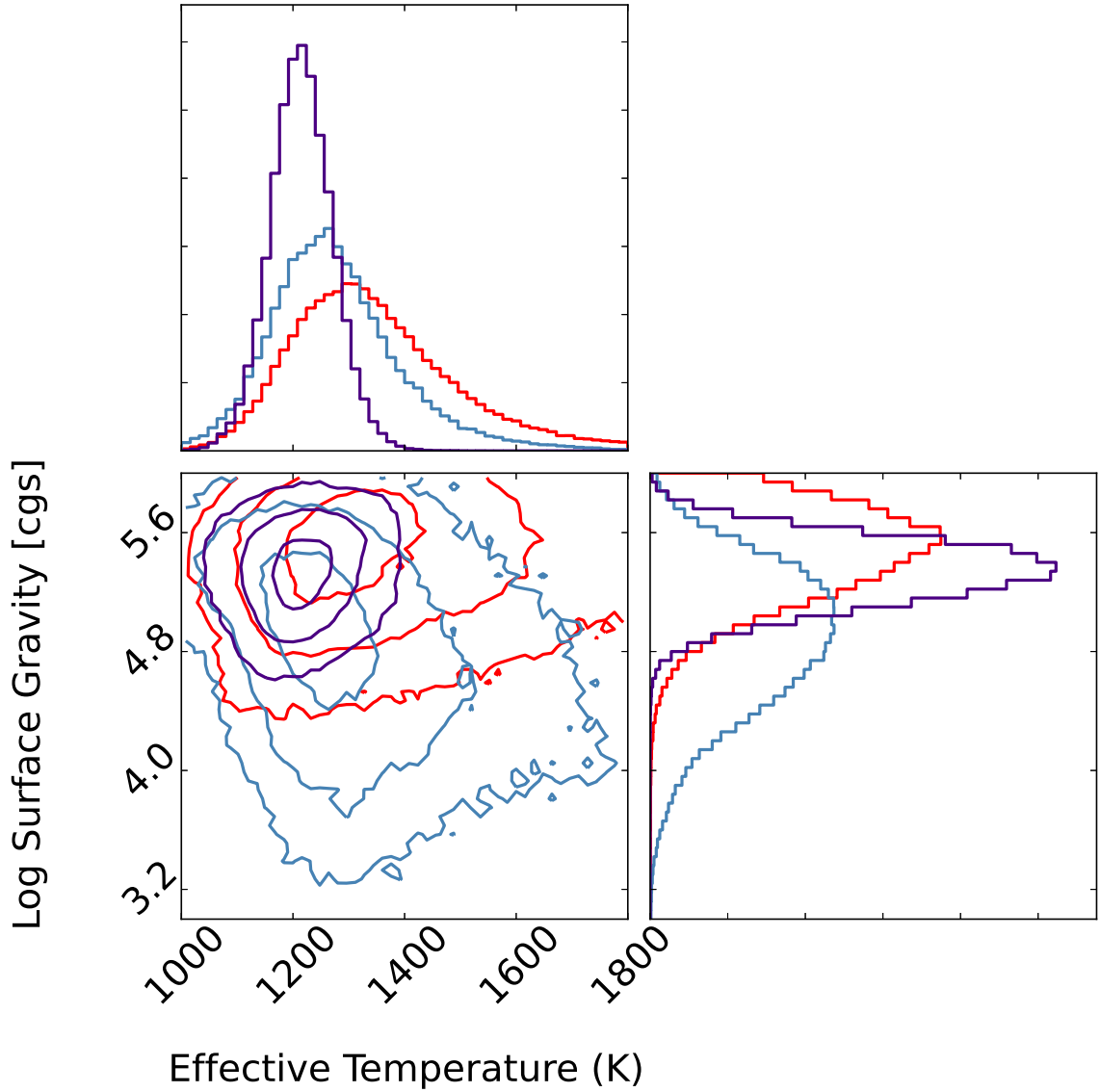}
  \caption{\label{mcmc_contours_T4.5}Same as Figure~\ref{mcmc_contours_M5} for the field T4.5 spectral template object. The results from different spectral bands are generally consistent, with long tails to higher effective temperatures for the individual $YJ$- and $H$-band spectra and to lower surface gravities for the $YJ$-band spectrum.
}
\end{figure}

Results from the spectral fitting routine described above are probability distributions with 10$^6$ values for each parameter. The probability distribution for fits to the simulated spectra from the field M5 spectral template are shown in Figure~\ref{mcmc_contours_M5}, for the field L1 in Figure~\ref{mcmc_contours_L1}, for the field L8 in Figure~\ref{mcmc_contours_L8}, and for the field T4.5 in Figure~\ref{mcmc_contours_T4.5}. In all figures the $YJH$ results are shown in purple, the $YJ$ results in blue, and the $H$ results in red. The lower-left panel shows 1-, 2-, and 3-$\sigma$ contours of the posterior distribution functions for both parameters, while the top and right panels shows the histograms for temperature and surface gravity, respectively, marginalized over the other parameter.

The probability distributions for the field M5 are presented in Figure~\ref{mcmc_contours_M5}. The $YJH$ and $YJ$-band spectra produce similar temperature results, but the $H$-band distribution has broad, flat peak with a long tail toward higher temperatures. None of the distributions shows a clear peak in surface gravity. None of the distributions shows a clear peak in surface gravity, with the $YJH$ distribution increasing toward unphysically high surface gravities and the $YJ$- and $H$-band distributions increasing slightly at values of log(g)$<$4.0 dex (cgs).

The probability distributions for the field L1 (Figure~\ref{mcmc_contours_L1}) are qualitatively similar to the M5 in temperature ($\sim$600~K cooler), with the $YJH$ and $YJ$-band spectra producing nearly overlapping symmetric and clearly peaked distributions, but the $H$-band distribution has a broad peak at hotter temperatures and low plateau at lower temperatures compared to the $YJH$ and $YJ$ peaks. In surface gravity, the $YJH$ and $H$-band distributions show clear peaks, while the $J$-band distribution is flat. Both the $YJH$ and $H$-band spectra show sharp decreases in probability at surface gravities $\ge$5.7, which are unphysically high values but allowed in the model grid.

The probability distributions for the field L8 are presented in Figure~\ref{mcmc_contours_L8}. The distribution for the $YJH$ spectrum is narrow and symmetric in both temperature and gravity. The $YJ$distribution has a broader, but still single-peaked, distribution in temperature but increases toward unphysical high values in surface temperature. The $H$-band distribution is broad in temperature, covering the entire range of both the $YJH$ and $YJ$ distributions in temperature. In gravity the $H$-band distribution is consistent with the $YJH$ distribution with a long, flat tail toward lower values.
 
The distributions for the T4.5 object (Figure~\ref{mcmc_contours_T4.5}) are the smoother and more symmetric in both parameters than those for the earlier spectral types. The peaks in the effective temperature distributions for each spectrum are offset by $\sim$50~K but are within the 1-$\sigma$ uncertainties of the other distributions. The surface gravity distributions are similarly offset, $H$-band to higher values and $YJ$-band to lower values, but are also consistent within the 1-$\sigma$ uncertainties. Both parameters are better constrained by complete $YJH$ spectra, with temperature more constrained by the $YJ$ fit and gravity more constrained by the $H$-band fit.

Comparing Figures~\ref{mcmc_contours_M5}--\ref{mcmc_contours_T4.5}, it becomes apparent that the low-resolution near-infrared spectra of very low mass objects are increasingly sensitive to temperature at later spectral types. Furthermore, the distributions from fits to the $YJ$-band are generally more consistent with those of the complete $YJH$ spectrum. The $YJ$ spectrum distribution for the L8 template object is an exception, but this is also the temperature regime where the {\it dusty} {\tt PHOENIX} models are not expected to accurately reproduce cool atmospheres. In surface gravity, the probability distributions also show tighter constraints for later spectral types, although these distributions sometimes extend into unphysical regimes of log(g)$\ge$5.6.

\subsection{Best Fit Spectra}
We adopt the best fit parameters of the 50\% quantile value for each parameter distribution, which corresponds to the peak of the histograms only for symmetric distributions. The uncertainties on the best fit parameters are defined by the 16\% and 84\% quantiles of the distributions, which correspond to 1-$\sigma$ width for a purely Gaussian posterior. Results are listed in Table~\ref{P1640results} and discussed below.

\begin{figure}
  \includegraphics[height=.75\textheight,angle=90,origin=c]{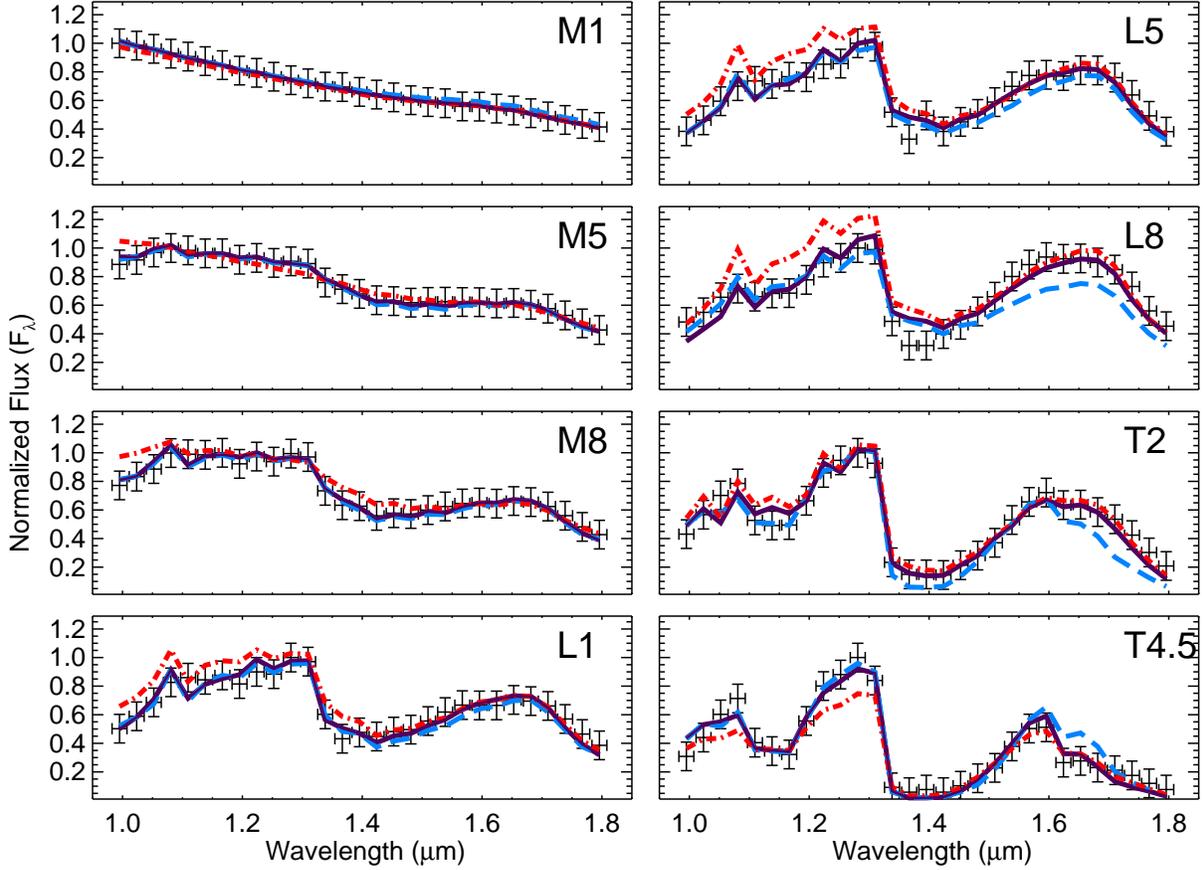}
  \caption{\label{P1640_bestfits}Best fit {\tt PHOENIX} model spectra for simulated P1640 spectra (gray error bars) of field M, L, and T dwarfs. Vertical error bars represent the constant noise value with SNR=10 at the peak flux, and horizontal errors bars represent the width of one wavelength channel. Purple lines represent the best fit model spectrum using all flux points in the simulated spectra except four points in the H$_2$O band 1.34--1.42~$\mu$m). Blue dashed lines represent the best fit model spectra using just the $\sim$$YJ$-band flux points (0.995--1.31~$\mu$m) and red dot-dashed lines, the $H$-band flux points (1.45--1.80~$\mu$m). M and L dwarf fits use the {\it dusty} version of the {\tt PHOENIX} models, and T dwarf fits use the {\it cond} version.
}
\end{figure}

\begin{figure}
  \includegraphics[height=.75\textheight,angle=90,origin=c]{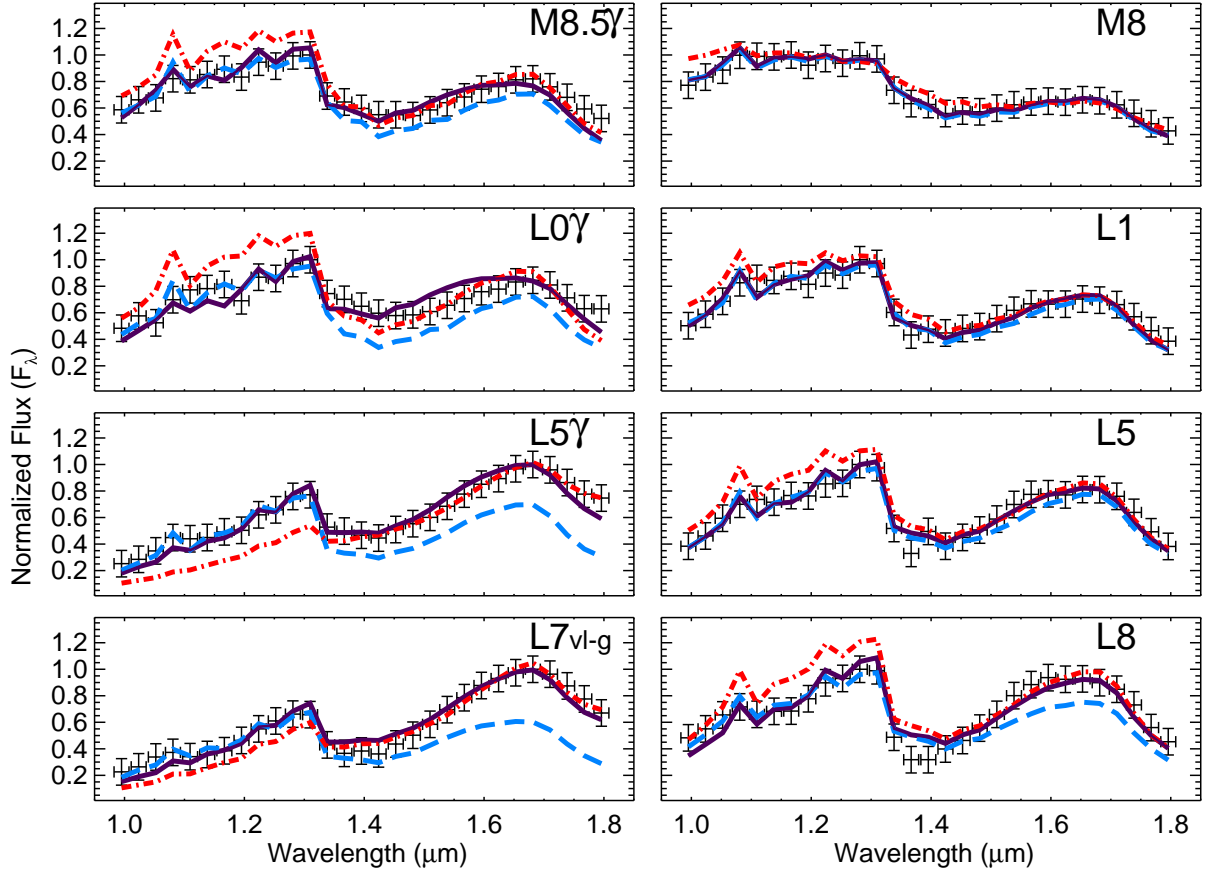}
  \caption{\label{P1640_bestfits_young}Best fit {\tt PHOENIX} {\it dusty} model spectra for simulated P1640 spectra (gray error bars) of young M and L spectral type objects (left) and their closest field-age counterparts in our sample (right). Vertical error bars represent the constant noise value with SNR=10 at the peak flux, and horizontal errors bars represent the width of one wavelength channel. Purple lines represent the best fit model spectrum using all flux points in the simulated spectra except four points in the H$_2$O band 1.34--1.42~$\mu$m). Blue dashed lines represent the best fit model spectra using just the $\sim$$YJ$-band flux points (0.995--1.31~$\mu$m) and red dot-dashed lines, the $H$-band flux points (1.45--1.80~$\mu$m). M and L dwarf fits use the {\it dusty} version of the {\tt PHOENIX} models.
}
\end{figure}

The effective temperature and surface gravity values of the best fit model spectra for each simulated P1640 spectrum are listed in Table~\ref{P1640results}, and the corresponding synthetic spectra are over-plotted with simulated spectra in Figures~\ref{P1640_bestfits} and \ref{P1640_bestfits_young}. Best fit parameters were determined via the two-step method described in Section~\ref{fits} above for four versions of the simulated spectra: complete $YJH$ spectra, ``trimmed'' $YJH$ spectra, and separate $YJ$ and $H$ spectra. The best fit results for the complete and trimmed $YJH$ spectra are effectively the same (see Section~\ref{trim} below) so only the complete $YJH$ results are reported and plotted, as solid purple lines, in Figures~\ref{P1640_bestfits} and \ref{P1640_bestfits_young}. Blue dashed lines represent the best fit synthetic spectrum with model parameters determined using only the $YJ$-band flux points (0.995--1.31~$\mu$m) and red dot-dashed lines, the $H$-band flux points (1.45--1.80~$\mu$m). All three best fit spectra are plotted over the complete $YJH$ wavelength range of the simulated spectra.

The best fit synthetic spectra from the $YJH$ fits reproduce the overall spectral shape and broad absorption features, within the assigned error bars of the simulated spectra, for all spectral types. The only marginal exceptions are 1--2 points in the the trough of the 1.4~$\mu$m \h2o absorption band for the field L5 and L8 objects, one point in the $Y$-band peak of the T4.5 dwarf, as well as two points around 1.4$\mu$m and the slope on either side of the $H$-band peak for the L0$\gamma$ object. This is remarkable especially considering the limiting-case dust treatments of the {\tt PHOENIX} {\it dusty} and {\it cond} models and the discrepancy of the model parameters with previous measurements from the literature (Sections~\ref{Teff} and \ref{logg}). Therefore we advise that even good fits should be treated with caution when interpreting fits to low-resolution near-infrared spectra for very low mass objects.

The fits to different subsets of the flux points, based on traditional filters separated by the telluric absorption band at 1.4~$\micron$, test the consistency of the best fit parameters from these different bands. For the M~dwarfs, all of the fits produce similar results, with the $H$-band fit slightly over predicting the first two or three $Y$-band flux points for the M5 and M8. For the L~dwarfs, the fits become less consistent, with the $H$-band parameters over predicting the flux for the $YJ$-band points and the $YJ$-band fit under predicting the flux at $H$-band, especially for the L8 object. The T~dwarfs are slightly better, with the only problems being the $YJ$ fits under-predicting the red end of the $H$-band for the T2 and slightly over predicting the flux for the T4.5, and the $H$-band fit slightly under-predicting the $Y$- and $J$-band peaks for the T4.5. 

The offsets between the synthetic spectra from $YJ$- or $H$-band fits and the rest of the simulated spectra are more significant for the young objects, particularly for the low gravity L~spectral types. Figure~\ref{P1640_bestfits_young} shows the best fit spectra for the young objects (left column) with the best fit spectra for field objects of similar spectral types (right column). For the hotter objects (M8.5$\gamma$ and L0$\gamma$) the $YJ$-band fits only slightly under-predict the $H$-band flux points (blue dashed line), while the $H$-band fits are more significantly different, especially for the L0$\gamma$ spectrum (red dot-dashed line). The single-band best fit spectra for the cooler objects (L5$\gamma$ and L7 {\sc vl-g}) always under-predict the flux in the other band, most substantially for the $YJ$-band fit in the $H$ band (blue dashed line).

Overall, the low-resolution fits to the complete $YJH$ spectra reproduce the simulated low resolution near-infrared spectra for all spectral types, despite using atmosphere models with limiting cases of dust treatment to produce the synthetic spectra. Fits to an individual bands result in significant discrepancies with flux points in the other band, suggesting that a broad wavelength coverage is best for reliable characterization of low mass companions via low resolution spectra.

\subsection{Effect of Trimming Spectra}
\label{trim}
The \h2o absorption band at $\sim$1.4~$\mu$m is typically low SNR and most likely to be affected by systematic errors induced by telluric calibration, which are difficult to quantify. These wavelength points are typically excluded from analysis for observed P1640 spectra \citep[e.g.,][]{Roberts12,Oppenheimer13}. Therefore, we fit ``trimmed" simulated P1640 spectra by excluding four flux points in the $\sim$1.4~$\mu$m \h2o absorption band to determine the effect of removing these flux values on the best fit model parameters. 

The resulting best fit parameters were identical within the uncertainties in both cases, with the largest differences being 15~K for the M5 object and 0.15--0.25~dex for the field L5 and L8 objects and the young M8.5$\gamma$ object. The temperature difference is 10 times smaller than the 1-$\sigma$ uncertainty defined by the 16\% and 84\% quantile values. The gravity differences are comparable to the 1-$\sigma$ uncertainties for those objects. Uncertainties derived from the probability distributions were also very similar, within 15~K and 0.04~dex log(g), except for the log(g) of the M8.5$\gamma$ object, with the 1-$\sigma$ quantile extending 1.4~dex lower in surface gravity for the trimmed spectrum. 

These results suggest that excluding low SNR flux measurements in the water band have little effect on the parameters of very low mass objects inferred from model fits to low resolution near-infrared spectra, but they might be marginally significant for hotter low-gravity objects. Results for the young objects are discussed in more detail in Section~\ref{young} below.

\subsection{Effective Temperature}
\label{Teff}
\begin{figure}
  \includegraphics[height=.90\textheight,angle=0,origin=c]{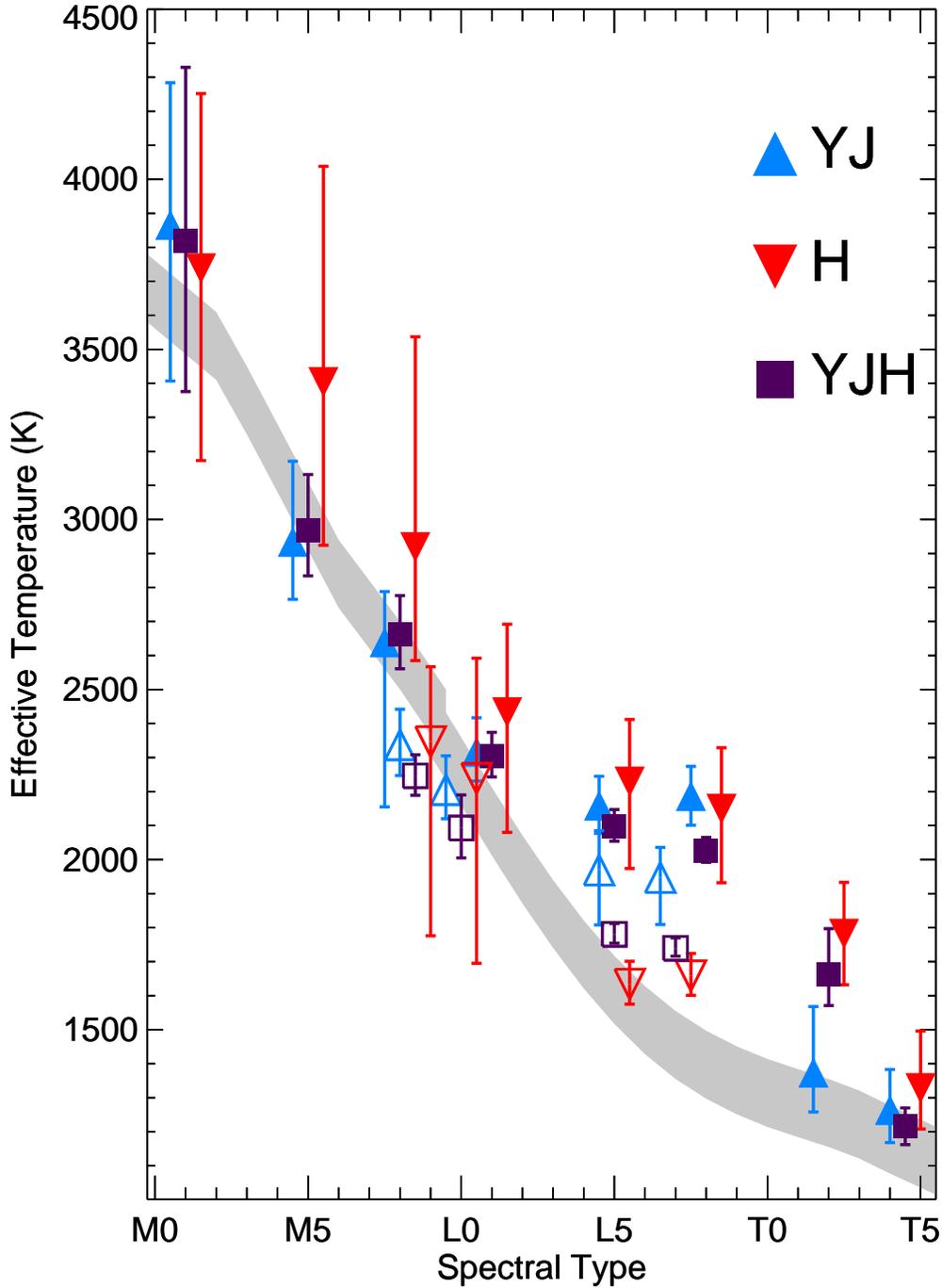}
  \caption{\label{P1640_teff}Best fit results in effective temperature versus spectral type for simulated P1640 data fit as $YJ$-band (blue upward triangles), $H$-band (red downward triangles) and $YJH$-band (purple squares) spectra. Field objects are represented by filled symbols and young objects, by open symbols. The gray filled region shows temperatures derived for objects with the same spectral types from \citet{Luhman99} for M dwarfs and \citet{Stephens09}  for L and T dwarfs with a range of $\pm$100~K.
}
\end{figure}

Figure~\ref{P1640_teff} shows the effective temperatures of the best fit synthetic spectra as a function of spectral type for the field objects (filled symbols) and young objects (open symbols) for each of the three versions of the spectra: $YJ$ (upward blue triangle), $H$ (downward red triangle), and complete $YJH$ (purple square). The triangle points are offset slightly in spectral type for clarity. The gray shaded region shows temperatures derived for the same spectral types from \citet{Luhman99} for M dwarfs and \citet{Stephens09} for L and T dwarfs with a range of $\pm$100~K.

For model fits to low-resolution near-infrared spectra to be considered reliable, they should be consistent with the physical parameters derived from other methods. Figure~\ref{P1640_teff} shows that the best fit results from low-resolution near-infrared spectra (symbols) match the empirically derived temperatures (gray shaded regions) to within the uncertainties for the field M dwarfs and the T4.5 dwarf, but the model fits result in hotter temperature for the field L5, L8, and T2 dwarfs, by as much as 500~K (for the L8 dwarf). This is the temperature regime in which dust is expected to condense from the photosphere, a dynamic and complex process that is not modeled in detail with the limiting cases of dust treatment in the {\tt PHOENIX} {\it dusty} and {\it cond} model atmospheres. 

The best fit effective temperatures from different segments of the spectra are generally consistent for a given object within the 1-$\sigma$ uncertainties, with notable exceptions of the $YJ$ fits for the L7 {\sc vl-g}, L8, and T2 objects. The $H$-band fits produce the highest temperatures for almost all of the field objects and lower temperatures for the young L dwarfs. 

Uncertainties in effective temperature are large (up to $\pm$500~K) for the earliest spectral types considered in our analysis, but these are still 10--20\% precision. Statistical uncertainties are as small as $\pm$30--50~K for the mid to late L dwarfs (field and young), although these values are also systematically hotter than the empirically derived temperatures, particularly for the field objects. As might be expected, the uncertainties are smallest for the complete $YJH$ spectra, but in some cases the uncertainties on individual bands are nearly as small, particularly for the $YJ$-band of the M8.5$\gamma$, L0$\gamma$, L5, and L7 objects and for the $H$-band of the L5$\gamma$, L7 {\sc vl-g}, and T2 objects.

The best fit effective temperature for known young objects are shown on Figure~\ref{P1640_teff} as open symbols. All young object spectral fits produce cooler temperature best fits than corresponding fits to spectra of field objects with similar spectral types. This is consistent with the idea that enhanced dust in cool, low-gravity atmospheres causes the near-infrared spectrum to appear more red (i.e., cooler) than would be predicted by the optically-defined spectral type (e.g., \citealt{Metchev06,Luhman07,Barman11b,Bowler13}). Indeed, direct empirical comparison of bolometric luminosities for young and field objects of the same spectral types suggest that, because they are similar but the young objects should have inflated radii, the young objects must have lower effective temperatures than their field-age spectral type counterparts (Filippazzo et al., in prep.).

\subsection{Surface Gravity}
\label{logg}
Substellar-mass objects never reach a stable main sequence; instead they cool, fade, and shrink for their entire lifetimes. Effective temperature and luminosity are degenerate with age and mass, but surface gravity is uniquely and significantly low ($<$5.0 dex cgs) for young objects ($<$100s~Myr). We include four known young substellar objects in our sample to test whether model fits are sensitive enough to surface gravity to distinguish young from old objects, or even to constrain age via comparison with evolutionary models, using low resolution near-infrared spectra. 

Figure~\ref{P1640_logg} shows the surface gravity of the best fit model as a function of spectral type with the same symbols as in Figure~\ref{P1640_teff} described in Section~\ref{Teff}. The shaded regions represent predictions for field (age$\ge$500~Myr, solid) and young (age 5--120~Myr, hatched) objects from evolutionary models: \citet{Siess00} for T$_{eff}$$>$3000~K, \citet{Baraffe02} and \citet{Chabrier00} [DUSTY00] for 3000$>$T$_{eff}$$>$1500~K, and \citet{Baraffe03} [COND0] for T$_{eff}$$<$1500~K. The hatched gray regions shows predictions from the DUSTY00 models for effective temperatures $\sim$2500--1500~K, which approximately corresponds  to the spectral types of young objects in our sample. 

Broadly, evolutionary models predict the surface gravity for field early-M~dwarfs to be in the lowest part of the shaded parameter range on Figure~\ref{P1640_logg}, i.e., log(g)=4.9~dex. The oldest late-L~dwarfs are predicted to peak at log(g)=5.4~dex, and the oldest, most massive T~dwarfs reach log(g)=5.5~dex. For substellar-mass objects, younger objects will have lower surface gravities as a result of their larger radii, but objects older than 500~Myr will have log(g)$>$4.9 for the entire range of temperatures considered, according to both the DUSTY00 and COND03 evolutionary models.

The best fit surface gravities are consistent with the values predicted by evolutionary models for field objects of all spectral types, within the relatively large 1-$\sigma$ uncertainties (from $\pm$0.2 to $\pm$0.9~dex). For the M and early-L~dwarfs, the $YJH$ best fit surface gravity is closer to the predicted value than the surface gravities from individual $YJ$ and $H$-band fits. For field L and T dwarfs, the best fit surface gravity is within the range of values predicted by the evolutionary models, except for the individual $YJ$- and $H$-bands for the L1 object. This is likely a result of the broad spectral features being increasingly shaped by surface gravity, in addition to temperature, for the coolest objects. It is perhaps surprising that such low resolution spectra are sensitive to surface gravity at all, but it has been shown that the relative strengths of broadband absorption like H$_2$O and collisionally-induced absorption from H$_2$ depend on surface gravity \citep[][Figure~6]{Rice11}.

The results from fits using different wavelength ranges for field objects are generally self-consistent to within the 1-$\sigma$ uncertainties. The largest discrepancy is for the field L5, where the individual $YJ$- and $H$-band fits indicate lower surface gravities ($\sim$5.0~dex cgs) and the complete $YJH$ fit indicates a higher surface gravity (5.35~dex cgs). The three fits indicate similar effective temperatures ($\sim$2100--2230~K), with the individual bands at the hotter end of the range. The results are all within the 1-$\sigma$ uncertainties for each parameter so are not likely to indicate any systematic trend.

The results for best fit surface gravity for the young objects are more discrepant from predictions of the evolutionary models and less consistent for fits from individual wavelength ranges. They are discussed in more detail in Section~\ref{young} below.

\subsection{Young Objects}
\label{young}
The primary targets of exoplanet direct-imaging campaigns are nearby young stars. Planetary-mass companions to young stars are more luminous than older objects because they are still contracting and radiating gravitational potential energy. Young brown dwarfs are similarly low mass and low surface gravity, but as free-floating objects they are amenable to a broader range of observations, both in wavelength coverage and spectral resolution. Therefore, young brown dwarfs are potentially important analogs to directly-imaged exoplanets (see, e.g., \citealt{Faherty13,Allers13}). 

We include four confirmed young brown dwarfs in our analysis in order to test whether model spectra reproduce the low resolution near infrared spectra of cool, low gravity objects and to compare the accuracy and consistency of best fit results for young and field objects with similar spectral types. The young template objects we use have optical or near-infrared spectral types of M8.5$\gamma$, L0$\gamma$, L5$\gamma$/L3 {\sc vl-g}, and L7 {\sc vl-g} (see \citealt{Cruz09} and \citealt{Allers13} for descriptions of the surface gravity suffixes) and medium resolution, high SNR spectra available in the literature (see Table~\ref{obs} for references). 

Figure~\ref{mcmc_compare} shows the posterior probability distribution functions for  young M8.5$\gamma$ and L5$\gamma$ objects and the field objects with the closest spectral types, with the $YJH$ results in purple, the $YJ$ results in blue, and the $H$ results in red. The scales on each axis are identical in effective temperature and log surface gravity in order to show systematic difference between results for objects on the same optically-defined spectral type but difference ages (field objects in top row and young objects in bottom row). For both young objects, the posterior distribution functions are centered on significantly cooler temperatures ($\Delta$T$_{eff}$=300--600~K) than for the comparable field object. For the young M8.5$\gamma$ object, the $YJH$ and $YJ$ fits are consistent in temperature, but the $H$-band fit is bimodal with the a significantly cooler component that is a very poor fit to the $YJ$ spectrum. For the young L5$\gamma$ object, the complete $YJH$ spectrum temperature distribution is consistent with the $YJ$ and $H$ distributions to within 1-$\sigma$. 

\begin{figure}
  \includegraphics[height=.75\textheight,angle=90,origin=c]{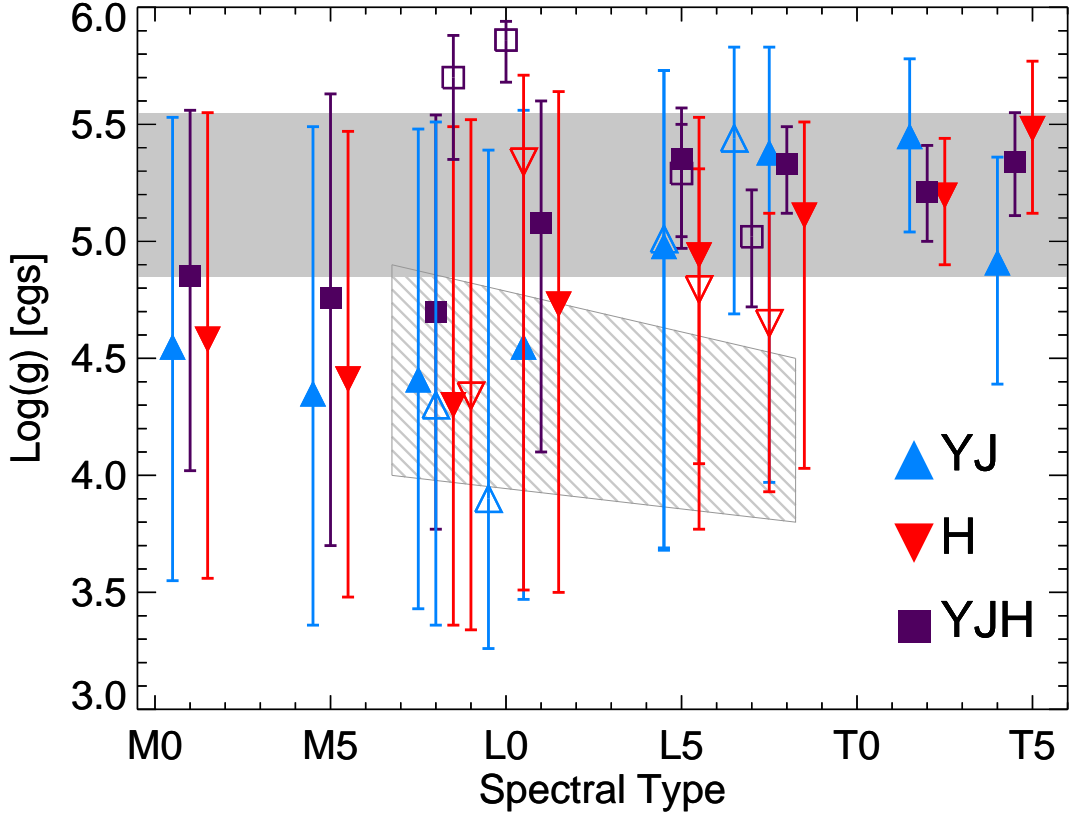}
  \caption{\label{P1640_logg}Best fit results in surface gravity versus spectral type for simulated P1640 data fit as $YJ$-band (blue upward triangles), $H$-band (red downward triangles) and $YJH$-band (purple squares) spectra. Field objects are represented by filled symbols and young objects, by open symbols. The solid shaded region shows the range of surface gravities for field objects (age~$\ge$~500~Myr) predicted by evolutionary models: \citet{Siess00} for T$_{eff}$$>$3000~K, \citet{Baraffe02} and \citet{Chabrier00} [DUSTY00] for 3000$>$T$_{eff}$$>$1500~K, and \citet{Baraffe03} [COND03] for T$_{eff}$$<$1500~K. The hatched gray regions shows predictions from the DUSTY00 models for effective temperatures $\sim$2500--1500~K, which corresponds approximately to the spectral types of young objects in our sample. The surface gravity results for field objects are closer to predictions from evolutionary models and more consistent between individual wavelength regions than the results for the young objects.
}
\end{figure}

\begin{figure}
  \includegraphics[height=.4\textheight,angle=0]{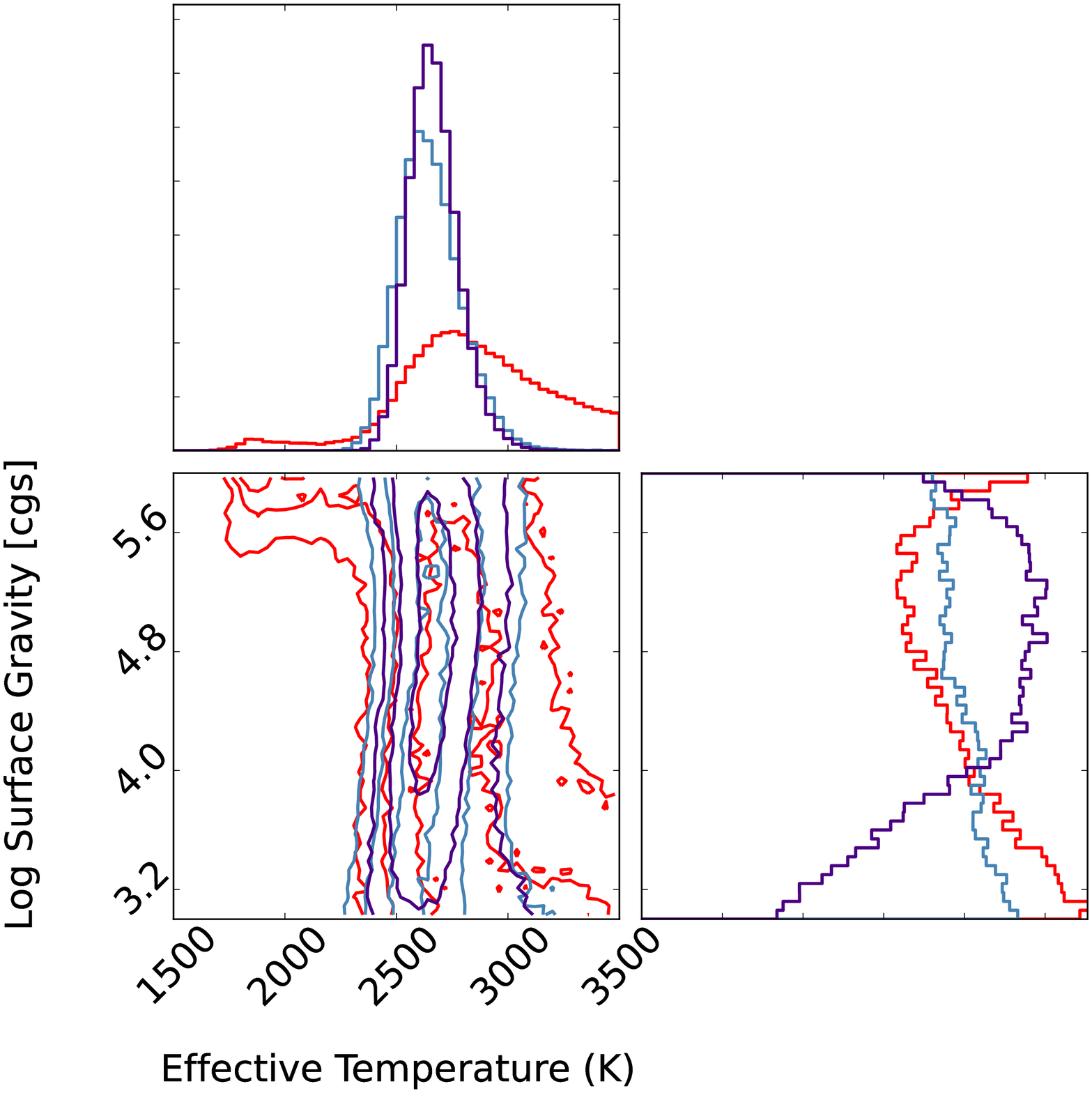} %M8
  \includegraphics[height=.4\textheight,angle=0]{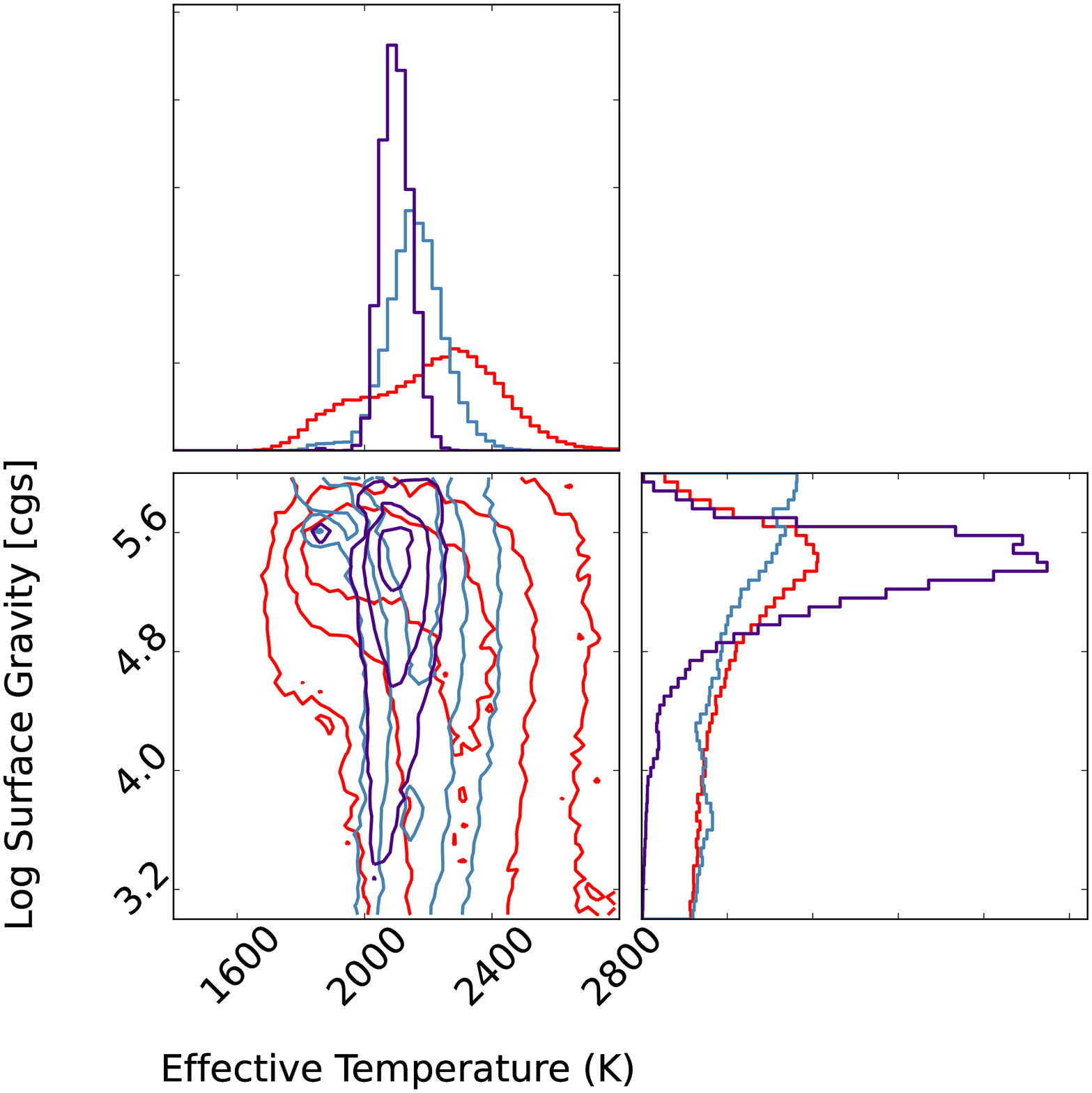} %L5
  \includegraphics[height=.4\textheight,angle=0]{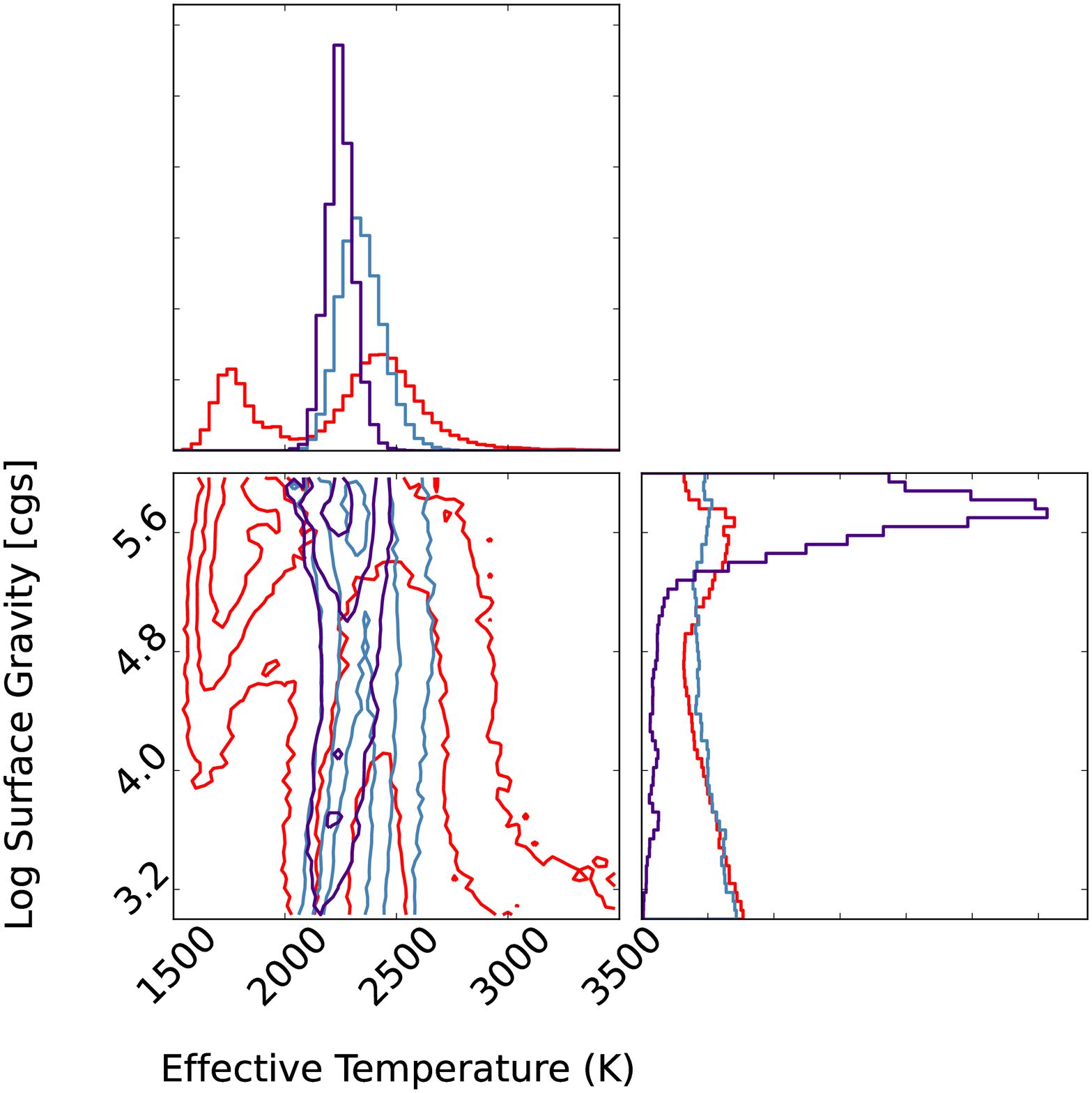} %M8.5gamma
  \includegraphics[height=.4\textheight,angle=0]{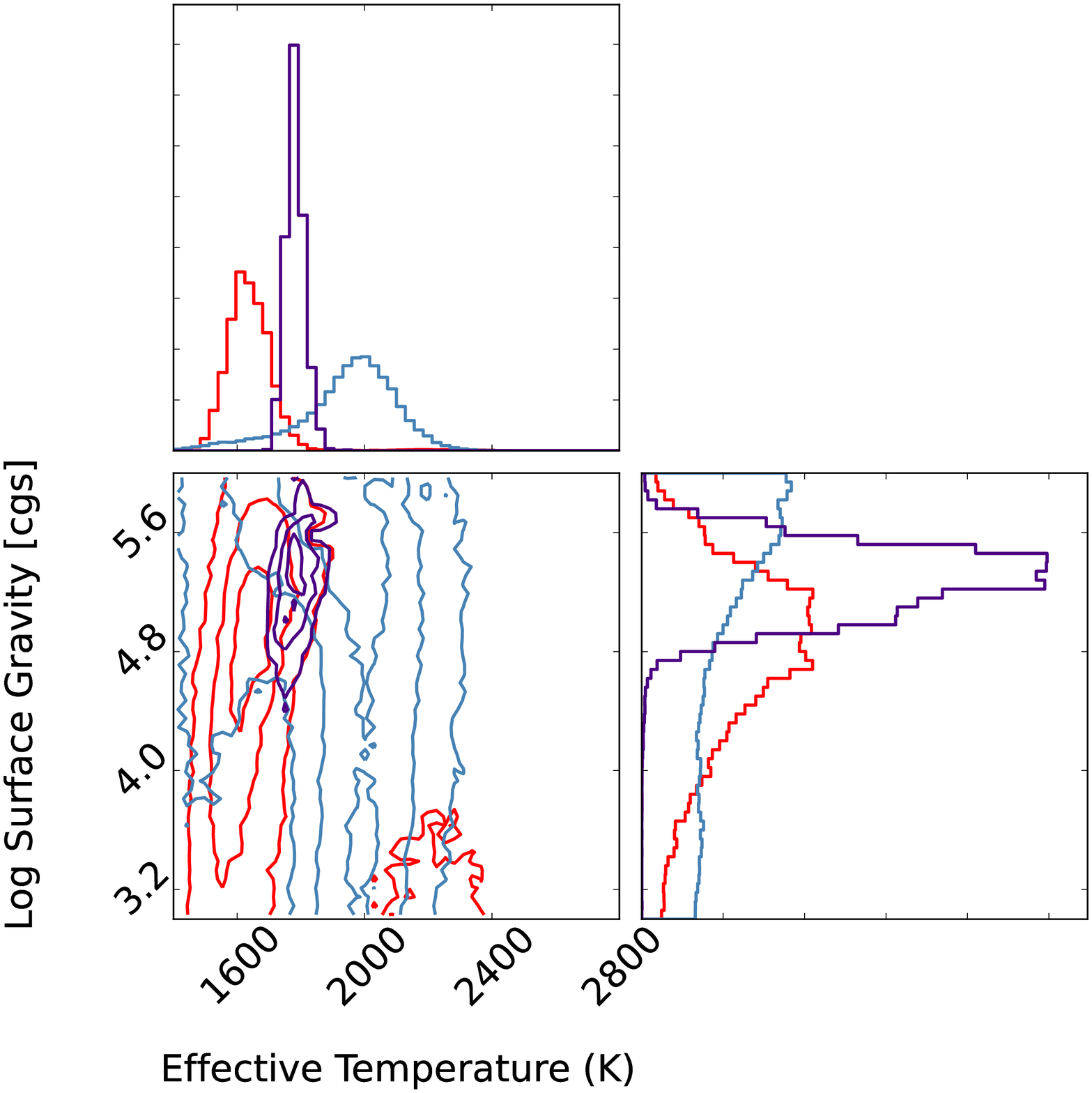} %L5gamma
  \caption{\label{mcmc_compare}MCMC results for young and field objects of approximately the same spectral type, $\sim$M8 objects on the left and L5 objects on the right, field objects on top and young objects on the bottom. Colors correspond to different spectral fits as in Figure~\ref{mcmc_contours_M5}. The peak of the distributions in effective temperature for the young objects are cooler than the corresponding peak for the field objects. Both young objects also have distributions that peak at higher surface gravities than predicted by evolutionary models using ages constrained by their likely membership in nearby young moving groups. See Section~\ref{young} for a complete discussion of the results.}
\end{figure}

\begin{figure}
  \includegraphics[height=.75\textheight,angle=0]{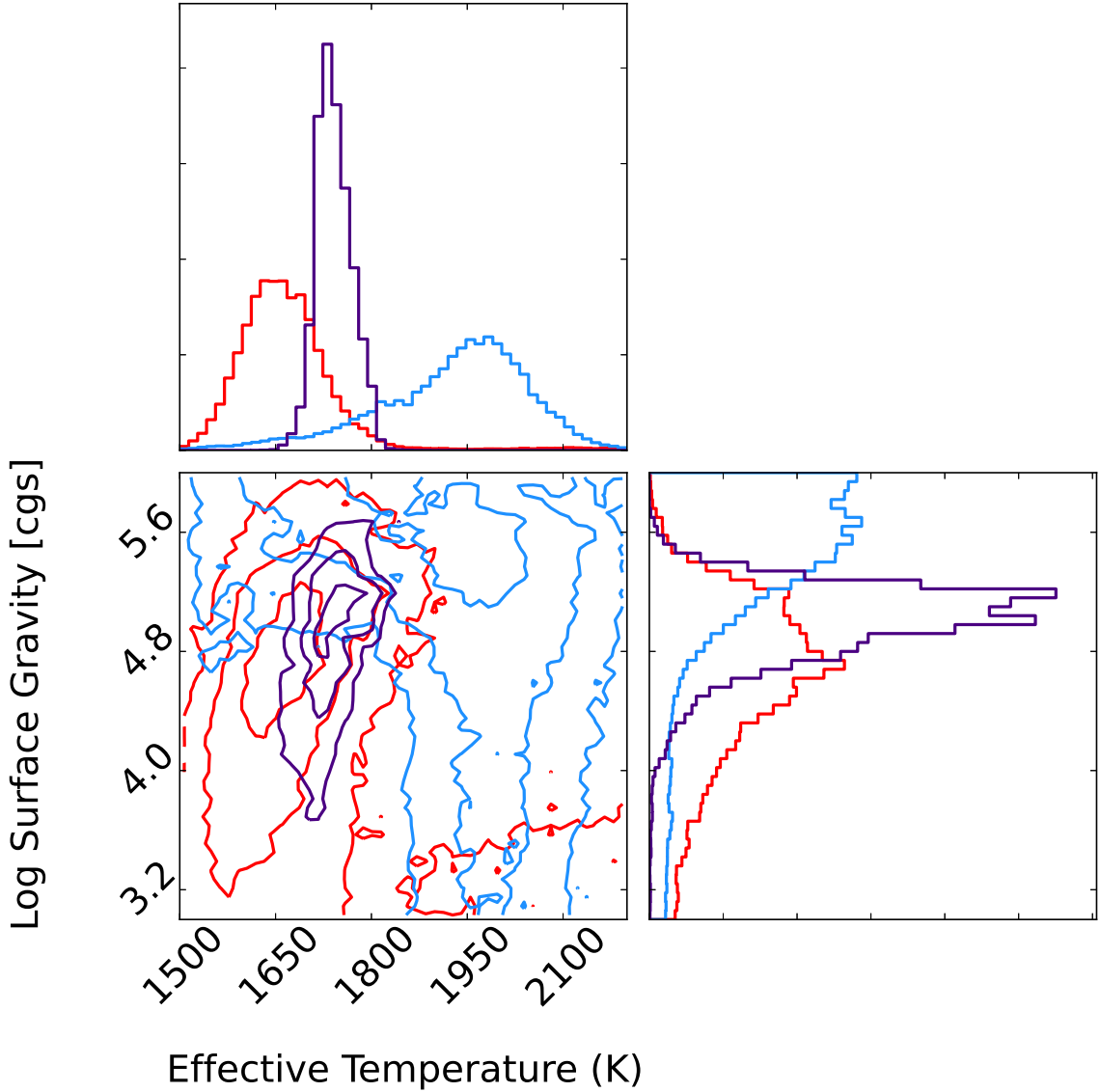}
  \caption{\label{mcmc_contours_L7}Same as Figure~\ref{mcmc_contours_M5} for the young L7 {\sc vl-g} object. The distributions from different spectral regions are substantially less consistent than for the field objects. For example, the 2-$\sigma$ contours of the $YJ$-band (blue) and $H$-band (red) distributions just barely overlap. While the $H$-band distribution peaks at a surface gravity predicted for a young, late-type object (see Figure~\ref{P1640_logg}), the $YJH$ and $YJ$-band distributions are peaked at higher surface gravities and have only a long tail toward lower values.  
}
\end{figure}

Figure~\ref{mcmc_contours_L7} shows the posterior probability distributions for model fits to the L7 {\sc vl-g} spectra. The distributions from fits to different spectral regions are substantially less consistent than for the field objects (Figures~\ref{mcmc_contours_M5}--\ref{mcmc_contours_T4.5}). For example, the 2-$\sigma$ contours of the $YJ$-band (blue) and $H$-band (red) distributions just barely overlap. The $H$-band distribution peaks at a surface gravity predicted for a young, late-type object (see Figure~\ref{P1640_logg}), but the $YJH$ and $YJ$-band distributions peak at higher surface gravities and have only a low tail toward lower values. This is further evidence that the {\tt PHOENIX} {\it dusty} models do not accurately reproduce the observed spectra of young, late-type objects \citep[e.g.,][]{Schmidt08,Mohanty10,Patience12}.   

Figure~\ref{P1640_bestfits_young} shows the simulated and best fit spectra for the young objects (left column) along with simulated and best fit spectra for field objects of similar spectral types (right column). The model spectra with the $YJH$ best fit parameters are plotted as solid purple lines, the $YJ$-band fit as dashed blue lines, and the $H$-band fit as dot-dashed red lines. For all four young objects, the complete (and trimmed) best fit $YJH$ spectrum reproduces the simulated spectrum to within the adopted $>$10\% uncertainties; however, the best fit parameters do not match predictions from evolutionary models, as discussed below.

The field M8 and young M8.5$\gamma$ have similar optically-determined spectral types but the $YJH$ spectrum of the young object has a more steeply sloped $YJ$ band, deeper H$_2$O absorption, and a higher, more triangular $H$-band peak.  Fits to the complete $YJH$ spectra are of similar quality, but fits to different wavelength ranges are less consistent for the young object than for its field-age counterpart. For the young object, the $YJ$-band fit slightly under-predicts the flux throughout the $H$-band, and even the $H$-band fit is slightly lower than the object's flux beyond 1.7~$\mu$m. The $H$-band fit over predicts the $YJ$-band flux even more, well outside of the adopted flux uncertainties. This is qualitatively similar to single-band results for the field L5 and L8 dwarfs, which are likely the dustiest of the field objects. 

The effective temperatures of the M8.5$\gamma$ best fit spectra are 400--600~K cooler than those for the field M8 (Figure~\ref{P1640_teff}). The best fit surface gravity for the young M8.5$\gamma$ is comparable to that of the and field object in $YJ$- and $H$-bands and match predictions from the DUSTY00 evolutionary models for 5--120~Myr objects (gray hatched region on Figure~\ref{P1640_logg}). The best fit surface gravity for the $YJH$ best fit spectrum is 1.0~dex (cgs) more than that of the field object, and is in fact higher than the maximum value predicted by evolutionary models for any substellar object \citep[e.g.,][]{Chabrier00,Baraffe02,Baraffe03}.

The complete $YJH$ fits to the simulated spectra of young L spectral types generally reproduce the flux points within the 1-$\sigma$ error bars, except for the longest-wavelength $H$-band point for three objects and several $J$- and $H$-band points for the L0$\gamma$ object. For the L0$\gamma$ object, the $YJH$ fit appears overall too flat at the blue end of the $H$ band, which corresponds to a high best fit surface gravity. The inconsistency between fits for individual bands is even more substantial for the young L objects than for field L dwarfs. For all three objects, the individual $YJ$ and $H$-band fits under predict the flux in the other band. The worst $H$-band fit in the $YJ$-band is for the L0$\gamma$, while the worst $YJ$-band fit in the $H$-band is for the L7 {\sc vl-g} object, which is the reddest of the young L dwarfs. 

All fits for the young Ls produce lower effective temperatures than for the field objects of similar spectral type, especially for the $YJH$ and $H$-band fits (Figure~\ref{P1640_teff}). The best fit surface gravities for the L0$\gamma$ are similar to that of the M8.5$\gamma$ in that the best fit parameters from the $YJH$ fits are too high, but the best fit parameters from the $YJ$- and $H$-band fits are comparable to predictions from the DUSTY00 evolutionary models for young objects. The surface gravity results for the later-type young L dwarfs are all higher than predicted by the evolutionary models, but not unphysical for field objects, and the $H$-band fits are closest to the predictions from evolutionary models (Figure~\ref{P1640_logg}), where the spectral shape is expected to be particularly sensitive to surface gravity because of decreased collisionally-induced absorption from H$_2$ relative to H$_2$O opacity (see e.g., \citealt{Rice11}., Figure~6).

Inconsistencies between model fits from individual bands, evident in both the probability distributions and in the best fit spectra, likely stem from the dust treatment in the {\it dusty} models, which do not incorporate enhanced dust content, particularly in small grains, expected for low surface gravity atmospheres (e.g., \citealt{Schmidt08,Marocco14}). These results support the growing consensus that additional opacity sources and more sophisticated dust and cloud treatments (e.g., \citealt{Witte09,Barman11a,Morley12,Morley14,Allard14}) are necessary for modeling the atmospheres of young brown dwarfs as well as of directly-imaged exoplanets.

These young, low-mass objects in particular illustrate the necessity of comparing observations in multiple bands in order to test the reliability of spectral characterization. Thus we caution against using results of model fits to low-resolution spectra to confirm or rule out youth of very low mass objects. Similar results for the P1640 spectrum of $\kappa$~Andromedae~B are described in \citet{Hinkley13}.

\subsection{Signal to Noise Ratio}
\label{snr}

The probability distributions, best fit parameters, and their uncertainties described above were calculated using the $\ge$10\% uncertainties assigned to the simulated spectra described in Section~\ref{P1640sim}. The noise level was selected such that the highest flux point of each simulated spectrum had SNR=10, which is expected to be typical for spectra of exoplanets from high contrast integral field spectrographs (e.g., \citealt{Oppenheimer13,Chilcote15}). The actual uncertainty on the binned flux points (i.e., the uncertainty on the observed flux points within the spectral range of each P1640 wavelength channel, added in quadrature) are significantly lower than 10\% because the original SpeX and GNIRS template spectra have significantly higher SNR and we have not adjusted the binned flux values at all. This begs the question of whether the results described above are truly representative of model fits to SNR$\le$10 spectra.

Therefore we use two Monte Carlo simulations to test the dependence of the best fit parameters and their uncertainties on the adopted SNR of the simulated spectrum. The first Monte Carlo test was applied to to all spectral types with the adopted (SNR$\le$10) uncertainties, and the second test used the M8, L5, and T4.5 simulated spectra to compare results from a range of different adopted uncertainty values. The complete $YJH$ simulated spectra were used for both SNR tests, and both tests compare simulated spectra to the calculated model spectral grid (i.e., 50~K intervals in effective temperature and 0.1 dex intervals in surface gravity, Figure~\ref{lowresmodels}), without the interpolation used in the MCMC fitting.

\subsubsection{Resampling versus real SNR}

The first SNR test, performed for all spectral types, resamples the simulated spectra by adding a random number to the binned flux values from the template spectra. The random number is drawn from a Gaussian distribution the width of the adopted noise level (SNR$\le$10), hereafter called a ``resampled" simulated spectrum. For 10$^4$ resampled simulated spectra, we calculate the goodness-of-fit to each calculated model spectrum, define the minimum goodness-of-fit value as the best fit parameters, and calculate the mean and standard deviation of the 10$^4$ best fit values. 

The mean best fit parameters for 10$^4$ resampled spectra of each template object are consistent with the $YJH$ results listed in Table~\ref{P1640results} to within one standard deviation. The largest discrepancies are for the M5 field object in surface gravity (different by $\sim$0.5~dex) and the L0$\gamma$ object in effective temperature (different by $\sim$60~K), both still within 1-$\sigma$ uncertainties of the MCMC results. Interestingly, the L0$\gamma$ object also has the worst ``best fit" $YJH$ spectrum (see Figure~\ref{P1640_bestfits_young}). The uncertainties were generally about the same or smaller (by up to a factor of 2) for the resampled spectrum fits, but this is possibly a consequence of using only the model grid instead of allowing for linear interpolation between calculated models.

\subsubsection{Dependence of Uncertainty on SNR}

\begin{figure}
  \includegraphics[height=.9\textheight,angle=0,origin=c]{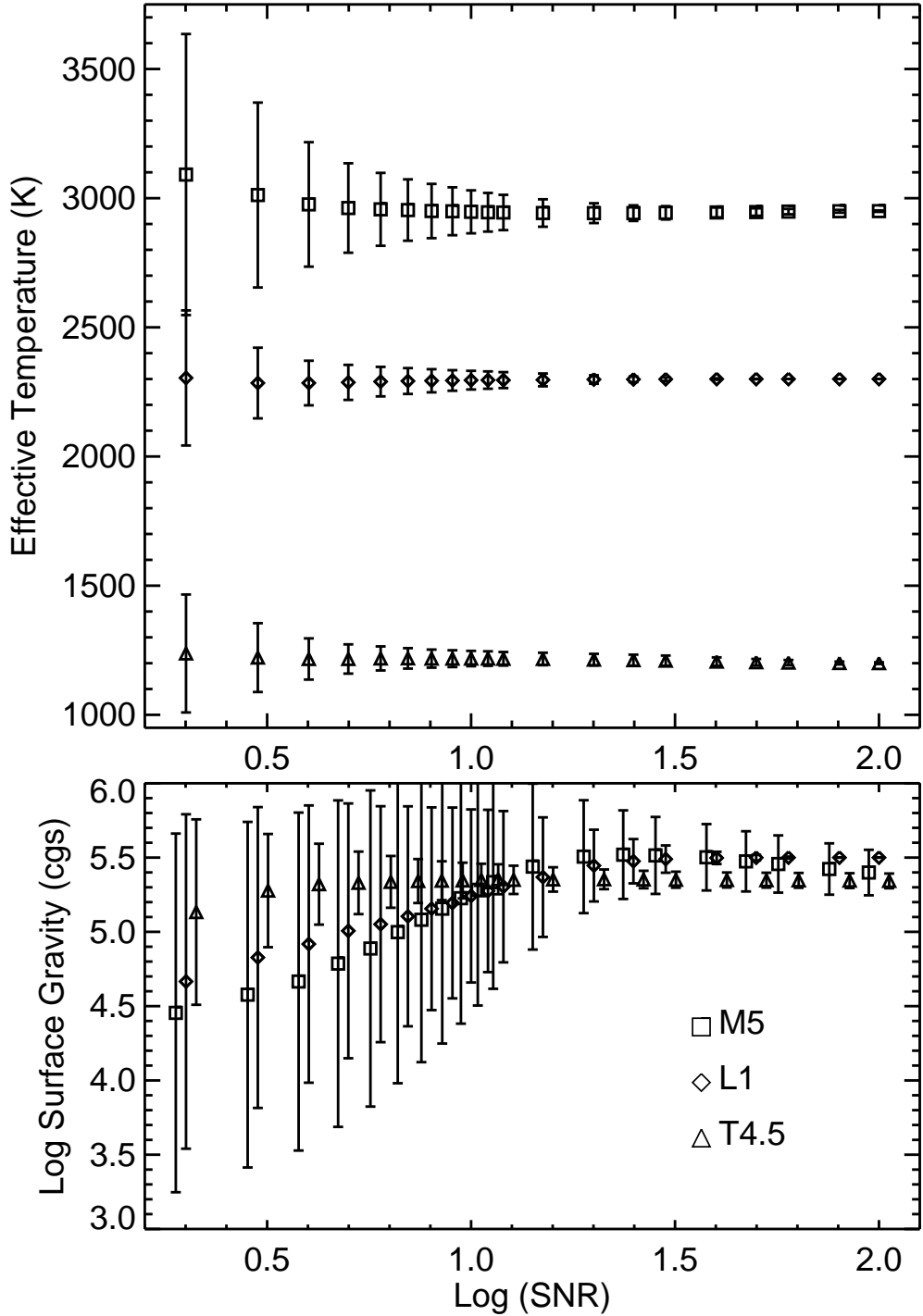}
  \caption{\label{snrs}Best fit effective temperatures (top) and surface gravities (bottom) for fits to the simulated P1640 spectra of the M5 (squares), L1 (diamonds), and T4.5 (triangles) dwarf templates with varied SNRs, as described in Section~\ref{snr}. Symbols on the bottom panel are offset slightly in log(SNR) for clarity. The error bars show the standard deviation of the distribution of best fit parameters for 10$^4$ resampled spectra using comparisons with the non-interpolated model grid. The results suggest that SNR$\ge$5 is required for temperature determination and SNR$\sim$10 is optimal for inferring surface gravity from best fit model spectra.
}
\end{figure}

For the second SNR test, we use the simulated spectra for the M5, L1, and T4.5 field objects to test the dependence of best fit parameters on SNR. As with the first test, the flux points are all resampled from within a Gaussian distribution the width of the noise on that flux point, this time for 20 values of maximum SNR, ranging from SNR=2 to 100. Each template spectrum is resampled 10$^4$ times for each SNR value, and the best fit model has the minimum goodness-of-fit value when compared to the resampled spectrum. The mean and standard deviation of the best fit parameters from 10$^4$ trials are plotted as a function of SNR in Figure~\ref{snrs}. 

The best fit parameters for the resampled simulated spectra approach T$_{eff}$=2950~K and log(g)=5.4 for the M5, T$_{eff}$=2300~K and log(g)=5.5 for the L1, and T$_{eff}$=1200~K and log(g)=5.3 for the T4.5. These values are consistent with the MCMC $YJH$ results within the 1-$\sigma$ uncertainties, but some are offset from the closet model grid point to the MCMC result, likely because of the fundamental difference between MCMC and purely Monte Carlo techniques and the use of interpolation versus the calculated model grid. 

By SNR$>$5, the mean temperatures for all three objects converge to within a few percent of the SNR=100 value, and the surface gravities converges to the SNR=100 values for SNR$>$10. The uncertainties in surface gravity remain large ($\sim$0.15~dex) for the M5 dwarf, likely because the low-resolution near-infrared spectra of M dwarfs are not as sensitive to gravity as the spectra of cooler objects are (i.e., Figure~\ref{mcmc_contours_M5}).

It should be noted that these tests assume a Gaussian noise model for the simulated IFS spectra, which may not be the most appropriate noise model for low SNR data. However, the Gaussian model is reasonable for a long chain of error sources with different characteristics, as is the case for high contrast IFS data. We expect the effects of the assumed error distribution to be smaller than the effect of the instrumental and the model systematics; therefore, we leave the evaluation of different noise models for a future paper. We also ignore instrumental systematic uncertainties mentioned in Sections \ref{P1640sim} and \ref{trim}. However, \citet{Oppenheimer13} tested the fidelity of their S4 spectral extraction method by injecting fake sources with a T4.5 spectrum into the data cubes obtained for HR~8799bcde and comparing the extracted spectra to the input spectrum. The average deviation from the input spectrum over all wavelengths was $\sim$2--9\%, except for the innermost planet HR~8799e, which was 15\%. Therefore we expect the instrumental systematics to be below the SNR$\sim$10 level we adopt for the simulated spectra. It should also be noted that the errors bars on Figure~\ref{snrs} are not representative of the actual uncertainty in the best fit parameters at high SNR, just the distribution of best fit parameters for resampled spectra, which will always approach a delta function at high SNR even for poor quality fits.

\section{\sc Conclusions}
\label{conclu}

We have developed a robust spectral fitting method and tested it for low-resolution ($R\sim$30 to 60) near-infrared ($YJH$) simulated spectra created from higher resolution observed template spectra of MLT dwarfs. We used the fitting method to explore the sensitivity of these spectra to effective temperature and surface gravity using the limiting-case dust treatments of the {\it dusty} and {\it cond} {\tt PHOENIX} atmosphere models for M/L dwarfs and T dwarfs, respectively. 

The {\tt PHOENIX} {\it dusty} and {\it cond} models reproduce the simulated spectra of both field and young objects when the entire wavelength regime is used in the fitting procedure, despite the limiting case dust treatments in these models. Best fit spectra determined using only one band ($YJ$ or $H$) are typically within the adopted uncertainties of the complete simulated spectra for the field dwarfs, with some exceptions for the L5, L8, and T dwarfs. However, the single-band best fits generally do not reproduce the spectrum in the other band for spectra of young objects, which leads us to recommend caution when using single bands of low-resolution spectra to characterize young, low-mass objects.

Our results indicate that low resolution near-infrared spectra are sensitive to temperature with precisions as good as $\pm$50--100~K for L and T dwarfs, while the constraints are much looser ($\pm$100--500) for M dwarfs. The low-resolution near-infrared spectra are increasingly sensitive to surface gravity at later spectral types, although with lower precision (up to $\sim$1~dex for hotter objects) and significant offsets from predictions of evolutionary models for the young objects. 

The complete and trimmed $YJH$ spectra produced similar results and uncertainties, indicating that losing a few flux points to poor sky subtraction, speckle suppression, or other reduction and extraction effects will not significantly affect the characterization of the companion. 

The best fit parameters and uncertainties can be considerably different for the $YJ$ and $H$ subsets of the complete $YJH$ spectrum, especially for temperature. Surface gravity results are generally consistent within the (large) uncertainties, but are discrepant from predictions of evolutionary models (see below). These results suggest that high contrast IFS observing campaigns should characterize detected companions with as broad a wavelength coverage as possible. In future analysis we will expand our simulated spectra into the $K$ band and test the sensitivity and accuracy of different filter combinations to model parameters.

The temperatures from the $YJ$ spectral fits are typically more consistent with the temperatures from the full $YJH$ fits and closer to temperatures derived for objects of the same spectral types using other methods with the notable exceptions of the young L5$\gamma$ and L7 {\sc vl-g} objects, for which the $H$-band fits are closer to complete $YJH$ results (Fig.~\ref{P1640_teff}). 

The best fit temperatures are most discrepant from literature values for the field L5 and L8 objects, likely the result of our use of the {\tt PHOENIX} {\it dusty} models. The best fit gravities are generally consistent with evolutionary models for the field objects but are significantly discrepant for the young objects. The complete $YJH$ spectra fits to young objects always produced high surface gravities. Even $H$-band spectra, which have been shown to be gravity sensitive at higher spectral resolutions \citep{Lucas01,Luhman05,Allers07,RFC10,Rice11}, only produced marginally consistent results for the young L dwarfs (see Figure~\ref{P1640_logg}). Thus we caution against using model fits to low-resolution near-infrared spectra to confirm or rule out youth of very low mass companions.

We advise that even good fits should be treated with caution when interpreting fits to low-resolution near-infrared spectra for very low mass objects. The results for simulated spectra from L-T transition objects and for young objects in particular highlight the need for testing the more sophisticated dust treatments now available in cool atmosphere models \citep[e.g.,][]{Morley12} and for comparing best fit parameters over a broad wavelength range.

We determine that a minimum of SNR$\sim$5 is required to reliably constrain the temperature of a low-mass companion and that SNR$\ge$10 is ideal for constraining surface gravity. We caution that these results are based on tests using objects with spectra that are reasonably well reproduced by the {\tt PHOENIX} {\it dusty} and {\it cond} models and may not be representative of results for all directly-imaged high contrast companions.

Future work will expand this analysis to simulated spectra for GPI and SPHERE, which cover $YJHK$ with several individual filters, and to include varying metallicity and dust treatments in the atmosphere models we consider. 

\section{\sc Acknowledgments}

The authors thank K.\@~Cruz, J.\@~Faherty, the BDNYC research group, \#AstroHackNY, and T.\@~Barman, D.\@~Brenner, I.\@~Crossfield, S.\@~Douglas, D.\@~Foreman-Mackey, A.\@~Kraus, L.\@~Hillenbrand, N.\@~Madhusudhan, and A.\@~Price-Whelan for useful discussions. We also thank the anonymous referee for both broad and detailed comments that helped us improve the manuscript.

This research was supported in part by the American Astronomical Society's Small Research Grant Program, NASA Astrophysics Data Analysis Program (ADAP) award 11-ADAP11-0169, and by the National Science Foundation under Grant No.\@~1211568. A portion of this work was supported by NASA Origins of the Solar System Grant No. NMO7100830/102190. A portion of the research in this paper was carried out at the Jet Propulsion Laboratory, California Institute of Technology, under a contract with the National Aeronautics and Space Administration (NASA) and was funded by internal Research and Technology Development funds. In addition, part of this work was performed under a contract with the California Institute of Technology (Caltech) funded by NASA through the Sagan Fellowship Program. The members of the Project 1640 team are also grateful for support from the Cordelia Corporation, Hilary and Ethel Lipsitz, the Vincent Astor Fund, Judy Vale, Andrew Goodwin, and an anonymous donor.  This research has made use of the IRTF Spectral Library, the SIMBAD database, operated at CDS, Strasbourg, France, and NASA's Astrophysics Data System. 

\bibliographystyle{apj}
\bibliography{ms}

\clearpage

\begin{deluxetable}{llllcccll}
\tabletypesize{\small}
\tablewidth{0pt} 
\tablecaption{\bf{IRTF/SpeX Spectra}
\label{obs} }
\tablehead{ 
\colhead{Spectral} &
\colhead{ Object} &
\colhead{ R.A.} &
\colhead{ Decl.} &
\colhead{ Ref.}  \\
\colhead{ Type} &
\colhead{ Name} &
\colhead{ (h, m, s)} &
\colhead{ (\degr,\arcmin,\arcsec)} &
\colhead{ }  }
\startdata
 \multicolumn{5}{l}{Field Templates} \\
\hline
M1 	& HD~42581  & 06 10 34.6 & $-$21 51 53  &1 \\ 
M5    & Gl 866ABC\tablenotemark{a} & 22 38 33.7  & $-$15 17 57 &1 \\
M8    & Gl 752B & 19 16 55.3 & +05 10 11  &1, 2 \\ 
L1     & 2MASS~J1439+1929 & 14 39 28.4  & +19 29 15  &1 \\ 
L5     & 2MASS~J1507$-$1627 & 15 07 47.7 & $-$16 27 39  &1 \\
L8     & DENIS~J0255$-$4700 & 02 55 03.6 & $-$47 00 51  &1\\ 
T2     & SDSS~J1254$-$0122 & 12 54 53.9 & $-$01 22 47 &1\\ 
T4.5  & 2MASS~J0559$-$1404 &05 59 19.1 & $-$14 04 49  &1\\ 
\hline
 \multicolumn{5}{l}{Young Templates} \\
\hline
M8.5$\gamma$    & 2MASS~J0608$-$2753 & 06 08 52.8 	& $-$27 53 58 & 3 \\ 
L0$\gamma$ & 2MASS~J0141$-$4633 & 01 41 58.23 & $-$46 33 57.4 & 4 \\ 
L5$\gamma$/L3 {\sc vl-g}  & 2MASS~J0355+1133 & 03 55 23.4 & +11 33 44 & 5, 6 \\ 
L7$\pm$1(IR) {\sc vl-g} & PSO~J318.5338$-$22.8603  	& 21 14 08.026 & $-$22 51 35.84 & 7 \\
\enddata
\tablerefs{(1)~\citet{Rayner09}, (2)~\citet{Cushing05}, (3)~\citet{RFC10}, (4)~\citet{Kirkpatrick06}, (5)~\citet{Faherty13}, (6)~\citet{Allers13}, (7)~\citet{Liu13}}
\tablenotetext{a}{Although Gl~866ABC is a triple system, \citet{RojasAyala12} find it to be close to Solar metallicity ([M/H]=0.05$\pm$0.12) with a temperature similar to single M5 dwarfs according to analysis of an unresolved R$\sim$2700 $K$-band spectrum.}
\end{deluxetable}

\begin{deluxetable}{lcccccc}

\tablewidth{0pt} 
\tablecaption{\bf{Spectral Fitting Results\tablenotemark{a} }
\label{P1640results} }
\tablehead{ 
\colhead{Spectral} &
\colhead{ }  &
\colhead{ T$_{eff}$ (K)} &
\colhead{ } & 
\colhead{ } &
\colhead{ log(g)} &
\colhead{ } \\
\colhead{Type} &
\colhead{ $YJH$} &
\colhead{ $YJ$} &
\colhead{ $H$} &
\colhead{ $YJH$} &
\colhead{ $YJ$} &
\colhead{ $H$}   }
\startdata
%		YJH					YJ					H					YJH			YJ				H
 \multicolumn{5}{l}{Field Objects} \\
\hline 
M1 	& 3820$^{+509}_{-444}$ & 3866$^{+418}_{-459}$ & 3738$^{+514}_{-565}$ & 4.85$^{+0.71}_{-0.83}$ & 4.55$^{+0.98}_{-1.00}$  & 4.58$^{+0.97}_{-1.02}$   \\ \\
M5    & 2969$^{+163}_{-135}$ & 2934$^{+237}_{-169}$ & 3404$^{+634}_{-480}$ & 4.76$^{+0.87}_{-1.06}$ & 4.35$^{+1.14}_{-0.99}$  & 4.41$^{+1.06}_{-0.93}$  \\ \\
M8    & 2661$^{+115}_{-100}$ & 2639$^{+149}_{-484}$ & 2919$^{+618}_{-334}$ & 4.70$^{+0.85}_{-0.93}$ & 4.41$^{+1.07}_{-0.98}$  & 4.30$^{+1.19}_{-0.94}$ \\ \\
L1     & 2305$^{+69}_{-62}$      & 2317$^{+100}_{-86}$   & 2433$^{+259}_{-353}$ & 5.08$^{+0.53}_{-0.98}$ & 4.55$^{+1.01}_{-1.08}$  & 4.73$^{+0.91}_{-1.23}$  \\ \\
L5     & 2098$^{+49}_{-44}$      & 2159$^{+86}_{-83}$      & 2229$^{+183}_{-355}$ & 5.35$^{+0.22}_{-0.38}$& 4.98$^{+0.75}_{-1.29}$  & 4.94$^{+0.59}_{-1.17}$  \\ \\
L8     & 2026$^{+39}_{-34}$      & 2187$^{+87}_{-86}$      & 2149$^{+180}_{-217}$ & 5.33$^{+0.16}_{-0.21}$ & 5.38$^{+0.45}_{-1.41}$ & 5.11$^{+0.40}_{-1.08}$ \\ \\
T2     & 1663$^{134}_{-92}$      & 1374$^{+194}_{-116}$ & 1782$^{+151}_{-150}$ & 5.21$^{+0.20}_{-0.21}$ & 5.45$^{+0.33}_{-0.41}$ & 5.20$^{+0.25}_{-0.30}$  \\ \\
T4.5  & 1215$^{+55}_{-53}$      & 1264$^{+119}_{-96}$   & 1328$^{+168}_{-120}$ & 5.34$^{+0.21}_{-0.23}$ & 4.91$^{+0.45}_{-0.52}$ & 5.48$^{+0.29}_{-0.36}$  \\
\hline
 \multicolumn{5}{l}{Young Objects} \\
\hline \\
M8.5$\gamma$ & 2246$^{+62}_{-57}$ & 2337$^{+105}_{-90}$ & 2347$^{+220}_{-571}$ & 5.70$^{+0.18}_{-0.35}$ & 4.30$^{+1.21}_{-0.94}$ & 4.34$^{+1.18}_{-1.00}$  \\ \\
L0$\gamma$    & 2093$^{+97}_{-88}$ & 2205$^{+100}_{-85}$ & 2239$^{+353}_{-544}$ & 5.89$^{+0.07}_{-0.10}$ & 4.03$^{+1.53}_{-0.76}$ & 4.74$^{+0.94}_{-1.37}$  \\ \\
L5$\gamma$     & 1780$^{+32}_{-26}$ & 1969$^{+116}_{-161}$ & 1634$^{+67}_{-59}$    & 5.29$^{+0.21}_{-0.27}$ & 5.01$^{+0.72}_{-1.33}$ & 4.80$^{+0.52}_{-0.75}$  \\ \\
L7 {\sc vl-g}	& 1739$^{+31}_{-23}$  & 1947$^{+89}_{-138}$ & 1659$^{+65}_{-58}$	  & 5.02$^{+0.20}_{-0.30}$ & 5.44$^{+0.39}_{-0.75}$ &4.65$^{+0.47}_{-0.72}$
\enddata
\tablenotetext{a}{Best fit parameters are 50\% quantile values, and quoted uncertainties are derived from 16\% and 84\% quantiles, equivalent to 1-$\sigma$ uncertainties for Gaussian posterior distributions (see Section~\ref{pdfs}).}
\end{deluxetable}

\end{document}